 \documentclass[journal]{IEEEtran}
\usepackage{cite}
\usepackage{amsmath,amssymb,amsfonts}
\usepackage{algorithm}  
\usepackage{algpseudocode}  
\usepackage{graphicx}
\usepackage{epstopdf}
\usepackage{textcomp}
\usepackage{xcolor}
\usepackage{bm}
\usepackage{enumerate}
\usepackage{amsmath}
\ifCLASSINFOpdf
\else
\fi
\begin{document}
	
\title{Multiple Intelligent Reflecting Surfaces Collaborative Wireless Localization System}
%\author{authors}
% \author{Ziheng~Zhang,~Wen~Chen,~\IEEEmembership{Senior~Member,~IEEE},~Zhendong~Li,~Xusheng~Zhu,~Qingqing~Wu,~\IEEEmembership{Senior~Member,~IEEE},\\~and~Jinhong~Yuan,~\IEEEmembership{Fellow,~IEEE\vspace{-2em}}
\author{Ziheng~Zhang,~Wen~Chen,~Qingqing~Wu,~Zhendong~Li, \\Xusheng~Zhu, Jingfeng Chen, and Nan~Cheng

% \thanks{This work is supported by National key project 2020YFB1807700, NSFC 62071296, Shanghai Kewei 22JC1404000. Wu's work is supported by NSFC 62371289 and NSFC 62331022.}

\thanks{Z. Zhang, W. Chen, Q. Wu, X. Zhu and J. Chen are with the Department of Electronic Engineering, Shanghai Jiao Tong University, Shanghai 200240, China (e-mail: zhangziheng@sjtu.edu.cn; wenchen@sjtu.edu.cn; xushengzhu@sjtu.edu.cn;
qingqingwu@sjtu.edu.cn;
laowu3917@sjtu.edu.cn).}
\thanks{Z. Li is with the School of Information and Communication Engineering, Xi'an Jiaotong University, Xi'an 710049, China (email: lizhendong@xjtu.edu.cn).}
\thanks{Nan Cheng is with the School of Telecommunications Engineering, Xidian University, Xi’an
710071, China (e-mail: nancheng@xidian.edu.cn).}
% <-this % stops a space
%\thanks{K. Wang is with the School of Information Science and Technology, ShanghaiTech University, Shanghai 201210, China, and also with the School of Communication and Electronic Engineering, East China Normal University, Shanghai 200241, China (e-mail: wangkl2@shanghaitech.edu.cn).}

%\thanks{This work is supported by ...}
\thanks{(\emph{Corresponding author: Wen Chen.})}}
\maketitle
\begin{abstract}
This paper studies a multiple intelligent reflecting surfaces (IRSs) collaborative localization system where multiple semi-passive IRSs are deployed in the network to locate one or more targets based on time-of-arrival. It is assumed that each semi-passive IRS is equipped with reflective elements and sensors, which are used to establish the line-of-sight links from the base station (BS) to multiple targets and process echo signals, respectively. Based on the above model, we derive the Fisher information matrix of the echo signal with respect to the time delay. By employing the chain rule and exploiting the geometric relationship between time delay and position, the Cramér-Rao bound (CRB) for estimating the target's Cartesian coordinate position is derived. Then, we propose a two-stage algorithmic framework to minimize CRB in single- and multi-target localization systems by joint optimizing active beamforming at BS, passive beamforming at multiple IRSs and IRS selection. For the single-target case, we derive the optimal closed-form solution for multiple IRSs coefficients design and propose a low-complexity algorithm based on alternating direction method of multipliers to obtain the optimal solution for active beaming design. For the multi-target case, alternating optimization is used to transform the original problem into two subproblems where semi-definite relaxation and successive convex approximation are applied to tackle the quadraticity and indefiniteness in the CRB expression, respectively. Finally, numerical simulation results validate the effectiveness of the proposed algorithm for multiple IRSs collaborative localization system compared to other benchmark schemes as well as the significant performance gains.

\end{abstract}


% \fi
\IEEEpeerreviewmaketitle

\section{Introduction}
The deep integration of sensing systems into future wireless networks is one of the prominent features of 6G \cite{9976205}. This integration mainly includes two aspects, one of which is the rise of environment-aware applications such as self-driving \cite{8809568}, unmanned aerial vehicle express \cite{9369901} and satellite navigation \cite{10355106}, and the other is reciprocity of communication and sensing, which means communication signals can be used for localization and imaging \cite{9785457} while the environmental information obtained through sensing can also improve the accuracy of channel estimation and reduce pilot overhead \cite{9765510}. However, providing high-precision sensing applications in various complex electromagnetic environments faces the following key challenges. First, high-precision localization services require high-frequency and large-bandwidth sensing signals; however, their propagation suffers from stronger free-space path loss as well as atmospheric absorption and attenuation, resulting in relatively short transmission distances \cite{9976205}. Then, it is difficult to establish a line-of-sight (LoS) link between the BS and the target due to various obstacles, especially in urban areas. Thus, solving the above challenges is the key to improving sensing performance in future wireless networks.

Intelligent reflecting surface (IRS) is a novel technology with the potential to address the aforementioned challenges. By forming a controllable electromagnetic field, it can reconstruct the channel and provide additional LoS links \cite{8811733}. Due to its low cost, ease of deployment, and advanced integration, it has been widely used in the new generation of wireless communication. Because of the many similarities between communication and sensing systems, there have been some recent research results on how the IRS is applied to assist sensing systems \cite{8910627}. Specifically, scholars have proposed three main IRS-assisted wireless sensing architectures: fully-passive IRS \cite{9361184,9456027},
%9133157,10440056
semi-passive IRS \cite{9370097,9340586} and IRS-self sensing systems \cite{9724202}. The main difference between these three lies in the relative position of the receiving antenna and transmitter with respect to the IRS. The performance of semi-passive IRS is optimal in most cases due to its lower path loss, making it one of the mainstream designs.

In addition to the design of basic architecture, more research has focused on system design,  performance analysis, and metrics selection. Detection probability is an important metric used in detection tasks in sensing. The optimal judgement threshold and closed expression for the detection probability for single target at different distances under the IRS-assisted sensing system were derived in \cite{9454375}. Because the expression of detection probability regarding the signal to interference plus noise ratio (SINR) ratio contains a Q-function, it is difficult to optimize as a metric for system design. The authors utilized the monotonicity property of detection probability regarding SINR and jointly optimize active and passive beamforming in multi-target scenarios \cite{9938373}. To reduce mutual interference between echo signals from different targets, three protocols, namely time division, code division, and hybrid division were also proposed in \cite{9938373}. Another important task in sensing is the estimation of parameters such as position, velocity, angle, etc. The Cramér-Rao bound (CRB) is used to evaluate the minimum mean square error between an estimated value and the true value under unbiased estimation. In \cite{10284917}, the authors derived the CRB for angle and distance estimation under optimal design for fully passive IRS and proposed a low-complexity estimation algorithm based on discrete Fourier transform. Furthermore, the authors compared the system design methods and CRB for point target and extended target; and analyzed the optimal allocation of reflection elements and sensors on semi-passive IRS \cite{peng2024semi}. \cite{10138058} considered the active and passive beamforming design under the integrated communication and sensing system and derived the CRB for direction-of-arrival estimation, and demonstrated the performance differences when the optimization objectives are SINR and CRB through simulation. However, the above 
research focuses on the derivation and optimization of CRB for of-arrival (DoA) estimation. In fact, it is difficult to uniquely determine the specific position of the target in space based solely on DoA. Research on estimation of target' specific position for IRS-assisted sensing systems needs to be conducted.

The above research on performance analysis and system optimization is based on a single IRS model. In fact,
due to the convenience of the IRS, there will be a large number of IRS deployed in various places in the future \cite{9475160}. In the next-generation wireless networks, the ubiquitous IRS can better serve communication and sensing applications through collaboration. First, multiple IRSs provide more observations from different angles. For multi-target sensing, multiple IRSs system can reduce the path loss of signal propagation. In \cite{9497709}, the authors jointly optimized the selection and beamforming among multiple IRSs to maximize energy efficiency and validated the advantages of multiple IRSs compared to single IRS and relay in simulation. The output probability of multiple IRS networks was analyzed in \cite{9205879} and the asymptotic expressions for the sum-rate was also derived. The above works ignore the secondary reflection among multiple IRSs, however, secondary reflection can be used to mitigate interference between multiple users \cite{9852985}. When IRSs are densely deployed in the network, there will be multiple links between the BS and the user through the reflections of different IRSs, the multi hop routing problem in multi IRS assisted communication was studied in \cite{9528043}.
% 9754265
However, very few studies have focused on multi-IRS-assisted sensing related systems. \cite{10506632} provided a two-stage protocol to serve integrated sensing and communication and pure communication separately. When the system locates the target, only the sensor operates in the semi-passive IRS, which is equivalent to the traditional sensing system. In \cite{10497119}, the authors considered multiple IRS serving multiple targets/users simultaneously, however, this service is independent, i.e., each user or target will only receive reflective signals from one IRS. Despite the research progress, the research on multiple IRSs collaborative localization has not been carried out.

Motivated by the above considerations, we study a multiple IRSs collaborative localization system where multiple semi-passive IRSs are deployed near the base station (BS) to locate one or more targets, as shown in Fig. 1. Our objective is to minimize the CRB for estimating the position of the target in the Cartesian coordinate system by jointly optimizing the active beamforming at the BS and IRS phase shifts. The main contributions of the paper are summarized as follows

\begin{itemize}
\item 
First, we propose a wireless localization system with multiple semi-passive IRSs collaboration, which has not been investigated in the literature. The reflective element reflects the sensing signal from the BS to the target while the sensor receives the echo signal from the target to complete the localization. To achieve orthogonality of multiplexed sensing signals, a joint design method for IRS selection and zero-forcing beamforming is proposed. Based on the above system design, we derive the Fisher information matrix (FIM) with respect to the delay of the multiplexed sensing signals passing through the target.
By means of the chain rule and the geometric relationship between delay and position, we derive the CRB for the estimation of the position of the target in the Cartesian coordinate.

\item Then, we propose a two-stage algorithm to minimize the derived CRB by jointly optimizing
the transmitting beamforming at the BS and the reflective beamforming at the multiple IRSs. Specifically, for the single-target case, we prove the monotonicity of the CRB with respect to the received energy of the target and derive the optimal closed-form solution for reflective beamforming. Then, we propose a low-complexity algorithm based on alternating direction method of multipliers (ADMM) to obtain the optimal solution for transmitting beaming design. For the multi-target case, we use duality to transform the max-min fairness problem into a quality-of-service (QoS) constraint problem. Due to the high coupling between IRSs coefficients and active beamforming coefficients, alternating optimization is used to transform the original problem into two subproblems. In these subproblems, semi-definite relaxation (SDR) and successive convex approximation (SCA) are applied to solve the quadraticity and indefiniteness in the CRB expression, respectively.

\item Finally, numerical simulations have verified that our proposed algorithm can achieve the best system performance under various parameter settings in single- and multi-target scenarios. It is also found that a small amount of semi passive IRSs can obtain effective multi-angle observations in space. When the performance improvement approaches saturation, the performance improvement brought by more semi-passive IRS is mainly due to the increase in sensors. Next, deploying more transmitting antennas than the number of IRSs at the BS allows the algorithm to run in only one stage, which reduces the computational complexity. Moreover, the simulation results also verify the proportional relationship between CRB and the number of reflective elements, sensors, and maximum transmitting power.
\end{itemize}

The rest of this paper is as follows. In Section II, we introduce the system model of multiple IRSs collaborative localization. Then, Section III derives the closed-form estimation CRB and formulates the optimization problem. Then, in Sections IV and V, we elaborate on the two-stage optimization algorithm for the single- and multi-target localization system, respectively. Section VI reveals the performance superiority of the proposed algorithm compared to other benchmarks and some useful insights. Finally, Section VII concludes this paper.

\textit{Notations:} Matrices, vectors and scalars
are represented by bold uppercase, bold lowercase and standard lowercase letters, respectively. For a complex-valued scalar $x$, $\left| {x} \right|$ denotes its absolute value. For a general matrix $\bf{A}$, $\text{rank}(\bf{A})$, ${\bf{A}}^H$ and ${{\left[\bf{A} \right]}_{i,j}}$ denote its rank, conjugate transpose and $\left( i,j \right)$-th element, respectively. For a square matrix $\bf{X}$, $\text{rank}(\bf{X})$, ${\bf{X}}^H$, and $\text{tr}(\bf{X})$ denote its rank, conjugate transpose, and trace. ${\bf{X}} \succeq 0$ denotes that $\bf{X}$ is a positive semidefinite matrix. ${\mathbb{C}^{M \times N}}$ represents the ${M \times N}$ dimensional complex matrix space. $\mathbb{E}\left( \cdot  \right)$ denotes the expectation operation. $\sim$ represents
“distributed as” and $\mathcal{C}\mathcal{N}\left( {\mathbf{x},\mathbf{R}} \right)$ represents
the distribution of a circularly symmetric complex Gaussian random vector with mean vector $\mathbf{x}$ and covariance matrix $\mathbf{R}$.
${{\left\| \cdot  \right\|}_{p}}$ represents the the $p$-norm of a vector. $\otimes $ represents the Kronecker product.

\section{System Model}
As shown in Fig. 1, the model of a multi-IRSs collaborative localization system is first introduced. We consider a IRSs-aided localization network where $K$ semi-passive IRSs are deployed to assist in the localization system from the BS with ${{N}_{T}}$ antennas to $Q$ targets. The direct link from BS to the target is blocked because of the obstacles around the BS. Each semi-passive IRS consists of two parts: $N$ passive reflecting elements and $M$ active sensors. In order to distinguish the index numbers of the passive refective elements and active sensors of the $k$-th semi-passive IRS, they are indexed by $k\in {{\mathcal{K}}}=\left\{ 1,\cdots K \right\}$ and $l\in {{\mathcal{K}}}=\left\{ 1,\cdots K \right\}$, respectively. The location of BS, $k$-th IRS, and $q$-th target are denoted by ${{\mathbf{l}}_{\text{BS}}}=\left( {{x}_{\text{BS}}},{{y}_{\text{BS}}},{{H}_{\text{BS}}} \right)$, ${{\mathbf{l}}_{\text{I},k}}=\left( {{x}_{\text{I},k}},{{y}_{\text{I},k}},{{H}_{\text{I},k}} \right)$ and ${{\mathbf{l}}_{\text{T},q}}=\left( {{x}_{\text{T},q}},{{y}_{\text{T},q}},0 \right)$, respectively.

\begin{figure}%[htbp]
\centerline{\includegraphics[width=8cm]{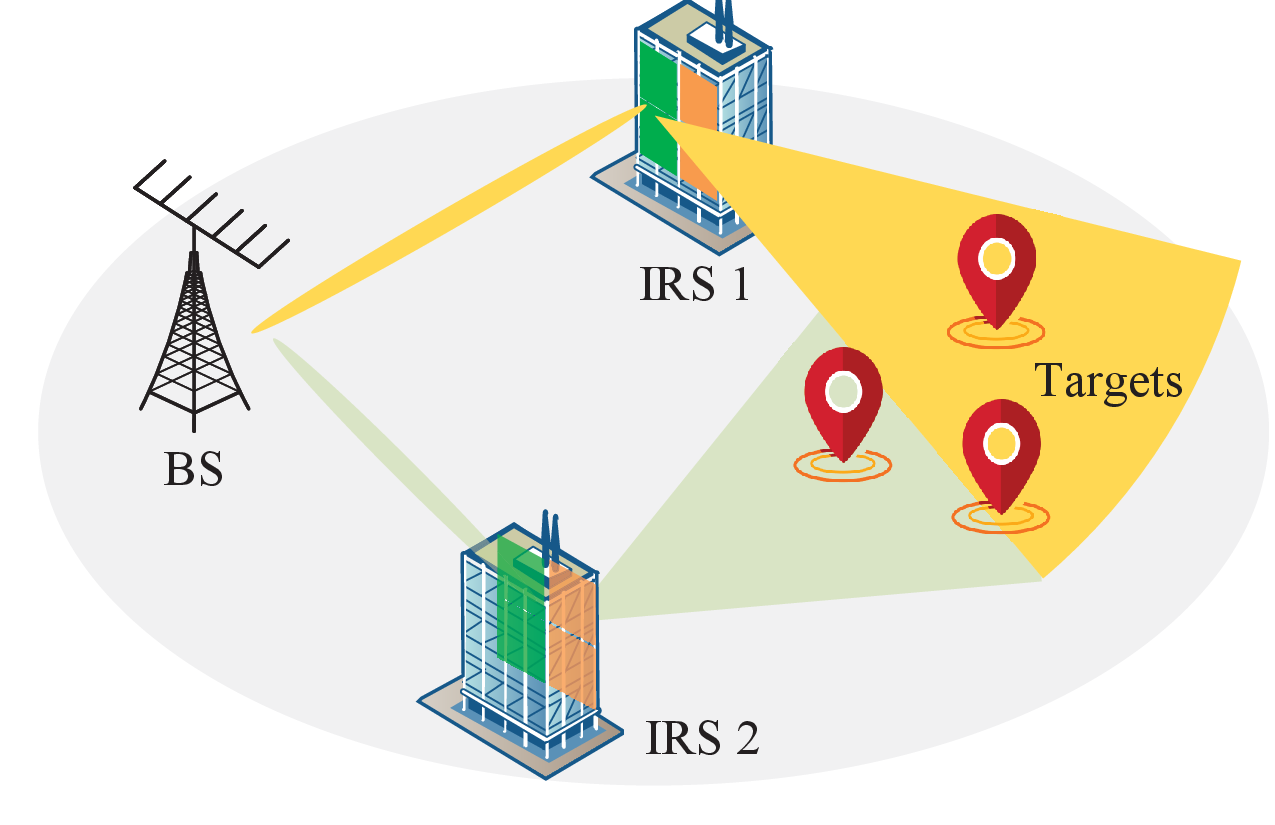}}
	\caption{Multiple IRSs collaborative localization system.}
	\label{Fig1}
\end{figure}

\subsection{Channel Model}
The relative direction of $k$-th IRS to
BS is defined by $\left\{ {{\theta }_{\text{B2I},k}},{{\varphi }_{\text{B2I},k}} \right\}$, where ${{\theta }_{\text{B2I},k}},{{\varphi }_{\text{B2I},k}}$ denote the azimuth and elevation angels from BS to $k$-th IRS, respectively. Here, we assume that the passive reflective elements and active sensing sensors are two uniform planar array, that is $N={{N}_{x}}\times {{N}_{z}}$ and $M={{M}_{x}}\times {{M}_{z}}$. ${{N}_{x}}$ and ${{N}_{z}}$ represent the number of elements in the horizontal and vertical directions of the passive reflective elements, respectively. ${{M}_{x}}$ and ${{M}_{z}}$ represent the number of elements in the horizontal and vertical directions of the active sensing sensors, respectively.
The channel power gain between BS and $k$-th IRS can be given by ${{\alpha }_{\text{B2I},k}}$ with ${{\alpha }_{\text{B2I},k}}=\sqrt{{{{\lambda }^{2}}}/{16{{\pi }^{2}}d_{\text{B2I},k}^{2}}}$, where $\lambda$ denotes the carrier wavelength and $d_{\text{B2I},k}$ denotes the distance between the BS and $k$-th IRS. The channel from BS to $k$-th IRS is modeled as
\begin{align}
{{\mathbf{H}}_{\text{B2I},k}}={{\alpha }_{\text{B2I},k}}{{\mathbf{a}}_{\text{I2B},k}}\mathbf{a}_{\text{B2I},k}^{H}\in {{\mathbb{C}}^{N\times N_T}},
\end{align}
where ${{\mathbf{a}}_{\text{I2B},k}}=\mathbf{a}\left( {{\Phi }_{\text{B2I},k}},{{N}_{x}} \right)\otimes \mathbf{a}\left( {{\Omega }_{\text{B2I},k}},{{N}_{z}} \right)$ represents the receive array response vector of the passive elements on $k$-IRS with ${{\Phi }_{\text{B2I},k}}=\sin \left( {{\varphi }_{\text{B2I},k}} \right)\cos \left( {{\phi }_{\text{B2I},k}} \right)$, ${{\Omega }_{\text{B2I},k}}=\cos \left( {{\varphi }_{\text{B2I},k}} \right)$, and ${{\mathbf{a}}_{\text{B2I},k}}=\mathbf{a}\left( {{\Omega }_{\text{B2I},k}},{{N}_{T}} \right)$ represents the transmit array response vector of the BS in the direction of the $k$-th IRS. $\mathbf{a}\left( \tilde{\theta },\tilde{N} \right)$ represents the steering vector where $\tilde{\theta }$ represents the phase difference between two adjacent elements or antennas, and $\tilde{N}$ represents the number of antennas transmitted or received, which is defined as
\begin{align}
\mathbf{a}\left( \tilde{\theta },\tilde{N} \right)={{\left[ 1,{{e}^{j2\pi \tilde{\theta }}},\cdots {{e}^{j2\pi \left( \tilde{N}-1 \right)\tilde{\theta }}} \right]}^{T}}\in {{\mathbb{C}}^{\tilde{N}\times 1}}.
\end{align}

It is assumed that $Q$ targets are point targets in the far field of each semi-passive IRS and they are indexed by $q\in {{\mathcal{Q}}}=\left\{ 1,\cdots Q \right\}$. The angle-of-departure from passive reflecting elements to target is equal to the angle-of-arrival from passive reflective elements of $k$-th IRS to the active sensing sensors of $l$-th IRS via $q$-th target
can be modeled as
\begin{align}
{{\mathbf{H}}_{\text{I2I},q,k,l}}={{\alpha }_{\text{I2I},q,k,l}}{{\mathbf{a}}_{\text{T2I},q,l}}\mathbf{a}_{\text{I2T},k,q}^{H}\in {{\mathbb{C}}^{{M}\times N}},
\end{align}
where ${{\mathbf{a}}_{\text{T2I},q,l}}=\mathbf{a}\left( {{\Phi }_{\text{T2I},q,l}},{{M}_{x}} \right)\otimes \mathbf{a}\left( {{\Omega }_{\text{T2I},q,l}},{{M}_{z}} \right)$ denotes the receive array response vector of the sensors on $l$-th IRS in the direction of the $q$-th target with ${{\Phi }_{\text{T2I},q,l}}=\sin \left( {{\varphi }_{\text{T2I},q,l}} \right)\cos \left( {{\theta }_{\text{T2I},q,l}} \right)$ and ${{\Omega }_{\text{I2T},q,l}}=\cos \left( {{\varphi }_{\text{I2T},q,l}} \right)$, ${{\mathbf{a}}_{\text{I2T},k,q}}=\mathbf{a}\left( {{\Phi }_{\text{I2T},k,q}},{{N}_{x}} \right)\otimes \mathbf{a}\left( {{\Omega }_{\text{I2T},k,q}},{{N}_{z}} \right)$ denotes the transmit array response vector of the passive elements on $k$-IRS in the direction of the $q$-th target. ${{\alpha }_{\text{I2I},q,k,l}}=\sqrt{{{{\lambda }^{2}}\kappa }/{64{{\pi }^{3}}d_{\text{I2T},k,q}^{2}d_{\text{T2I},q,l}^{2}}}$ denotes the path loss, where $\kappa $ denotes the radar cross section, ${{d}_{\text{I2T},k,q}}$ and ${{d}_{\text{T2I},q,l}}$ denotes the distance from $q$-th target to $k$-th IRS and $l$-th IRS, respectively.
\subsection{Localization Model}
Then, we introduce the model and principles of localization for semi-passive IRS system. Let ${{\mathbf{\Theta }}_{k}}=\text{diag}\left( {{\boldsymbol{\theta }}_{k}} \right)=\text{diag}\left( {{\alpha }_{k,1}}{{e}^{j{{\beta }_{{k},1}}}},\cdots,{{\alpha }_{k,N}}{{e}^{j{{\beta }_{k,N}}}}\right)$ denotes the reflection coefficient matrix of $k$-th IRS, where ${{\alpha }_{k,n}}$ and ${{\beta }_{k,n}}$ denotes the amplitude and phase reflection coefficient of the $n$-th element of $k$-th IRS, respectively. Each reflective elements of the IRS will reflect signals from BS to $Q$ targets while each sensors of IRS will receive echo signals from $Q$ targets. For $l$-th IRS, the received echo signal at its sensors can be denoted as
\begin{align}
  & {{\mathbf{r}}_{\text{IRS},l}}\left( t \right)=\sum\limits_{q=1}^{Q}{\sum\limits_{k=1}^{K}{{{\mathbf{H}}_{\text{I2I},q,k,l}}{{\mathbf{\Theta }}_{k}}{{\mathbf{H}}_{\text{B2I},k}}}} \nonumber\\ 
 & \cdot {{\mathbf{w}}_{k}}{{s}_{k}}\left( t-{{\tau }_{\text{B2I},k}}-{{\tau }_{\text{I2T},k,q}}-{{\tau }_{\text{T2I},q,l}} \right)+{{\mathbf{n}}_{l}},
\end{align}
where ${\mathbf{n}_{l}}$ represents Gaussian white noise at the sensors in $l$-th IRS and ${{\bf{n}}_l} \sim \mathcal{C}\mathcal{N}\left( {0,\sigma _l^2{{\bf{I}}_{{M}}}} \right)$; ${{\tau }_{\text{B2I},k}}$ represents the time-of-arrival (TOA) from BS to $k$-th IRS. Because the location of the BS and IRS is fixed, ${{\tau }_{\text{B2I},k}}$ is known and useless for localization. ${{\tau }_{\text{T2I},q,l}}$ represent the TOA from $q$-th target to $l$-th IRS. ${{\tau }_{\text{I2T},k,q}}$ and ${{\tau }_{\text{T2I},q,l}}$ are unknown and key parameters in the localization system. Based on the elliptical localization method, ${{\tau }_{\text{I2T},k,q}}$ and ${{\tau }_{\text{T2I},q,l}}$ are viewed as a whole, so we define ${{\tau }_{q,k,l}}\triangleq {{\tau }_{\text{I2T},k,q}}+{{\tau }_{\text{T2I},q,l}}$. The relationship between ${{\tau }_{q,k,l}}$ and location of the $q$-th target can be expressed as
\begin{align}\label{location}
  c{{\tau }_{q,k,l}}={{\left\| {{\mathbf{l}}_{\text{I},k}}-{{\mathbf{l}}_{\text{T},q}} \right\|}_{2}}+{{\left\| {{\mathbf{l}}_{\text{I},l}}-{{\mathbf{l}}_{\text{T},q}} \right\|}_{2}}. 
\end{align}

\section{CRB Performance For Targets' Estimation And Problem Formulation}
\subsection{Cramér-Rao Bound}
CRB provides a lower bound of variance of any unbiased estimator, given the location vector to be estimated of $Q$ targets $\mathbf{u}={{\left[ {{\mathbf{u}}_{1}},\cdots ,{{\mathbf{u}}_{Q}} \right]}^{T}}={{\left[ {{x}_{\text{T},1}},{{y}_{\text{T},1}},\cdots ,{{x}_{\text{T},Q}},{{y}_{\text{T},Q}} \right]}^{T}}\in {{\mathbb{R}}^{2Q}}$. For each element in $\mathbf{u}$, the variance of the unbiased estimator ${{\widehat{\mathbf{u}}}_{i}}$ satisfies
\begin{align}
\operatorname{var}\left( {{\widehat{\mathbf{u}}}_{i}} \right)\ge {{\left[ {{\mathbf{C}}_{CRB}}\left( \mathbf{u} \right) \right]}_{i,i}}={{\left[ {{\mathbf{F}}^{-1}}\left( \mathbf{u} \right) \right]}_{i,i}},i=1,2,\cdots 2Q,
\end{align}
where ${{\mathbf{C}}_{CRB}}\left( \mathbf{u} \right)$ and ${{\mathbf{F}}^{-1}}\left( \mathbf{u} \right)$ denote the CRB matrix and FIM of $\mathbf{u}$ to be estimated, respectively. However, it’s difficult to calculate the FIM with respect to $\mathbf{u}$ directly. Accordingly, we choose to calculate the FIM with respect to $\boldsymbol{\tau}\in {{\mathbb{R}}^{Q\cdot {{K}^{2}}}}$ first, where $\boldsymbol{\tau}=\left[ {{\tau }_{1,1,1}},\cdots ,{{\tau }_{1,K,K}},\cdots ,{{\tau }_{Q,K,K}} \right]$ and apply the chain rule to drive the $\mathbf{F}\left( \mathbf{u} \right)$. Specifically, $\mathbf{F}\left( \mathbf{u} \right)$ can be represented as
\begin{align}\label{fu}
\mathbf{F}\left( \mathbf{u} \right)=\frac{\partial \boldsymbol{\tau}}{\partial \mathbf{u}}\mathbf{F}\left( \boldsymbol{\tau} \right){{\left( \frac{\partial \boldsymbol{\tau}}{\partial \mathbf{u}} \right)}^{T}}\in {{\mathbb{R}}^{2Q\times 2Q}}.
\end{align}
Applying the theory of matrix partitioning, $\mathbf{F}\left( \boldsymbol{\tau} \right)$ has the following form
\begin{align}\label{fv}
\mathbf{F}\left( \boldsymbol{\tau} \right)=\left[ \begin{matrix}
   {{\mathbf{F}}_{1,1}} & {{\mathbf{F}}_{1,2}} & \cdots  & {{\mathbf{F}}_{1,Q}}  \\
   \mathbf{F}_{1,2}^{H} & {{\mathbf{F}}_{2,2}} & \cdots  & {{\mathbf{F}}_{2,Q}}  \\
   \vdots  & \vdots  & \ddots  & \vdots   \\
   \mathbf{F}_{1,G}^{H} & \mathbf{F}_{1,G-1}^{H} & \cdots  & {{\mathbf{F}}_{Q,Q}}  \\
\end{matrix} \right]\in {{\mathbb{R}}^{G{{K}^{2}}\times G{{K}^{2}}}},
\end{align}
where the $\left({\left( {{k}_{1}}-1 \right)K+{{l}_{1}},\left( {{k}_{2}}-1 \right)K+{{l}_{2}}}\right)$-th
element of submatrix ${{\mathbf{F}}_{{{q}_{1}},{{q}_{2}}}}$ is given by 
\begin{align}
  & {{\left[ {{\mathbf{F}}_{{{q}_{1}},{{q}_{2}}}} \right]}_{\left( {{k}_{1}}-1 \right)K+{{l}_{1}},\left( {{k}_{2}}-1 \right)K+{{l}_{2}}}}={{\mathbb{E}}_{{{\mathbf{r}}_{\text{IRS}}}|\boldsymbol{\tau }}}\left[ \frac{{{\partial }^{2}}\log p\left( {{\mathbf{r}}_{\text{IRS}}}|\boldsymbol{\tau } \right)}{\partial {{\tau }_{{{q}_{1}},{{k}_{1}},{{l}_{1}}}}\partial {{\tau }_{{{q}_{2}},{{k}_{2}},{{l}_{2}}}}} \right], \nonumber\\ 
 & \left( {{k}_{1}},{{k}_{2}},{{l}_{1}},{{l}_{2}}\in {\mathcal{K}},{{q}_{1}},{{q}_{2}}\in {\mathcal{Q}} \right) 
\end{align}
where ${{\mathbf{r}}_\text{IRS}}=\left[ {{\mathbf{r}}_{\text{IRS},1}},{{\mathbf{r}}_{\text{IRS},2}},\cdots {{\mathbf{r}}_{\text{IRS},K}} \right]$ represents the echo signal observations collected by active sensors of $K$ IRSs, $p\left( {{\mathbf{r}}_\text{IRS}}|\boldsymbol{\tau}\right)$ is the joint probability density function of ${{\mathbf{r}}_{\text{IRS}}}$ conditioned on $\boldsymbol{\tau}$ and can be given by \cite{5466526}
\begin{align}
p\left( {{\mathbf{r}}_{\text{IRS}}}|\boldsymbol{\tau} \right)=\exp \left( -\frac{1}{\sigma _{l}^{2}}\sum\limits_{l=1}^{K}{\int_{T}{{{\left| {{\mathbf{r}}_{\text{IRS},l}}\left( t \right)-\overline{{{\mathbf{r}}_{\text{IRS},l}}}\left( t \right) \right|}^{2}}dt}} \right),
\end{align}
where $\overline{{{\mathbf{r}}_{\text{IRS},l}}}\left( t \right)$ denotes the deterministic component of the received echo signal. 

It is assumed that $G$ targets all are isolated. Specifically, arbitrarily different two targets ${{q}_{1}}$ and ${{q}_{2}}$ are said to be separable over the $k$-$l$ path, which means the signal reflected from elements of $k$-th IRS and received at sensors of $l$-th IRS. If the radar effective pulse width is ${{\tau }_{r}}$, the following inequalities hold \cite{8964303}
\begin{align}
\left| {{\tau }_{{{q}_{1}},{{k}_{1}},{{l}_{1}}}}-{{\tau }_{{{q}_{2}},{{k}_{2}},{{l}_{2}}}} \right|>{{\tau }_{r}},\forall {{q}_{1}}\ne {{q}_{2}},{{k}_{1}},{{k}_{2}},{{l}_{1}},{{l}_{2}}.
\end{align}
Based on the above assumptions, echo signals from different targets are separable in the time domain and do not interfere with each other. Therefore, based on (9) and (10), ${{\left[ {{\mathbf{F}}_{{{q}_{1}},{{q}_{2}}}} \right]}_{\left( {{k}_{1}}-1 \right)K+{{l}_{1}},\left( {{k}_{2}}-1 \right)K+{{l}_{2}}}}$ is denoted as (\ref{fnm1}) and the detail is shown in Appendix A,
% \newcounter{my6}
\begin{figure*}[!t]
	\normalsize
\begin{equation}\label{fnm1} 
{{\left[ {{\mathbf{F}}_{{{q}_{1}},{{q}_{2}}}} \right]}_{\left( {{k}_{1}}-1 \right)K+{{l}_{1}},\left( {{k}_{2}}-1 \right)K+{{l}_{2}}}}=\left\{ \begin{aligned}
  & 0,{{l}_{1}}\ne {{l}_{2}}\text{ or }{{q}_{1}}\ne {{q}_{2}},\\ 
 & \frac{1}{\sigma _{l}^{2}}\operatorname{Re}\int\limits_{T}{\left[ {{\mathbf{H}}_{{{q}_{1}},{{l}_{1}},k}}\mathbf{H}_{{{q}_{2}},{{l}_{2}},k}^{H}\sum\limits_{k=1}^{K}{\mathbf{w}_{k}^{H}\dot{s}_{k}^{H}\left( t-{{\tau }_{{{q}_{1}},{{l}_{1}},k}} \right)}\sum\limits_{k=1}^{K}{{{\mathbf{w}}_{k}}{{{\dot{s}}}_{k}}\left( t-{{\tau }_{{{q}_{2}},{{l}_{2}},k}} \right)} \right]dt},\text{others},\\ 
\end{aligned} \right.
\end{equation}
	\hrulefill
	\vspace*{1pt}
\end{figure*}
where ${{\mathbf{H}}_{q,k,l}}={{\mathbf{H}}_{\text{I2I},q,k,l}}{{\mathbf{\Theta }}_{k}}{{\mathbf{H}}_{\text{B2I},k}}$ and ${{\dot{s}}_{k}}\left( t-{{\tau }_{q,k,l}} \right)=\frac{\partial {{s}_{k}}\left( t-{{\tau }_{q,k,l}} \right)}{\partial {{\tau }_{q,k,l}}}$.
Although the echo signals from different targets do not interfere, the reflected signals from passive units of different IRSs reflected by the same target can interfere with each other when they are received at the sensor. To ensure that sensors can distinguish different delay signals of the same target,
reflected signals from different IRS must be orthogonal, i.e. 
\begin{align}
{{\mathbf{H}}_{\text{I2I},{{k}_{1}},l}}{{\mathbf{\Theta }}_{{k}_{1}}}{{\mathbf{H}}_{\text{B2I},{{k}_{1}}}}{{\mathbf{w}}_{{k}_{2}}}=\mathbf{0},{{k}_{1}}\ne {{k}_{2}}.
\end{align}
To meet above constraints, we can design the beam assignment of the BS to achieve orthogonality of the reflected signals or zeros the amplitudes of IRSs, i.e.,
\begin{align}\label{zero}
{{\mathbf{a}}^{H}}\left( {{\Omega }_{\text{B2I},{{k}_{1}}}},{{N}_{T}} \right){{\mathbf{w}}_{{k}_{2}}}=0,{{k}_{1}}\ne {{k}_{2}}\text{ or }{{\mathbf{\Theta }}_{{k}_{1}}}={{\mathbf{0}}_{K\times K}}.
\end{align}
In the subsequent system design, we propose a two-stage algorithm to minimize the CRB by using above two methods simultaneously. Based (\ref{zero}), ${{\left[ {{\mathbf{F}}_{{{q}_{1}},{{q}_{2}}}} \right]}_{\left( {{k}_{1}}-1 \right)K+{{l}_{2}},\left( {{k}_{2}}-1 \right)K+{{l}_{2}}}}$ is denoted as
\begin{align}
& {{\left[ {{\mathbf{F}}_{{{q}_{1}},{{q}_{2}}}} \right]}_{\left( {{k}_{1}}-1 \right)K+{{l}_{1}},\left( {{k}_{2}}-1 \right)K+{{l}_{2}}}}= \nonumber\\
&\left\{ \begin{aligned}
& \frac{\eta}{\sigma _{l}^{2}}\left\| {{\mathbf{H}}_{{{q}_{1}},{{k}_{1}},l}}{{\mathbf{w}}_{{k}_{1}}} \right\|_{2}^{2},{{q}_{1}}={{q}_{2}},{{k}_{1}}={{k}_{2}} \text{ and }{{l}_{1}}={{l}_{2}},\\ 
 & 0,\text{others},\\ 
\end{aligned} \right.
\end{align}
where ${\eta}={\int_{T}{{{{\dot{s}}}_{{k}_{1}}}\left( t-{{\tau }_{{{q}_{1}},{{k}_{1}},{{l}_{1}}}} \right)\dot{s}_{{k}_{1}}^{*}\left( t-{{\tau }_{{{q}_{1}},{{k}_{1}},{{l}_{1}}}} \right)dt}}$.
Then, we calculate the relationship between time delay and position $\frac{\partial \boldsymbol{\tau}}{\partial \mathbf{u}}$, which can be given by matrix partitioning as follows
\begin{align}\label{vu}
\frac{\partial \boldsymbol{\tau}}{\partial \mathbf{u}}=\left[ \begin{matrix}
   \frac{\partial \boldsymbol{\tau}}{\partial {{\mathbf{u}}_{1}}} & \cdots  & {{\mathbf{0}}_{2\times {{K}^{2}}}}  \\
   \vdots  & \ddots  & \vdots   \\
   {{\mathbf{0}}_{2\times {{K}^{2}}}} & \cdots  & \frac{\partial \boldsymbol{\tau}}{\partial {{\mathbf{u}}_{Q}}}  \\
\end{matrix} \right]\in {{\mathbb{R}}^{2Q\times {{K}^{2}}}},
\end{align}
where $\frac{\partial \boldsymbol{\tau}}{\partial {{\mathbf{u}}_{q}}}$ can be given by
\begin{align}\label{vuq}
\frac{\partial \boldsymbol{\tau}}{\partial {{\mathbf{u}}_{q}}}=\left[ \begin{matrix}
\frac{\partial {{\tau }_{1,1,1}}}{\partial {{x}_{\text{T},q}}} & \cdots  & \frac{\partial {{\tau }_{1,K,K}}}{\partial {{x}_{\text{T},q}}}  \\
\frac{\partial {{\tau }_{1,1,1}}}{\partial {{y}_{\text{T},q}}} & \cdots  & \frac{\partial {{\tau }_{1,K,K}}}{\partial {{y}_{\text{T},q}}}  \\
\end{matrix} \right]\in {{\mathbb{R}}^{2\times {{K}^{2}}}}.
\end{align}
Based on (\ref{location}), The derivative of the time delay with respect to the coordinates $\frac{\partial {{\tau }_{q,k,l}}}{\partial {{x}_{\text{T},q}}}$
and
$\frac{\partial {{\tau }_{q,k,l}}}{\partial {{y}_{\text{T},q}}}$
can be denoted as (\ref{taux}) and (\ref{tauy}), respectively. In order to more clearly represent the geometric relationship between the target and multiple IRSs, we explain the meaning of the angles mentioned in Fig. \ref{angle}.
\begin{figure}%[htbp]
\centerline{\includegraphics[width=8cm]{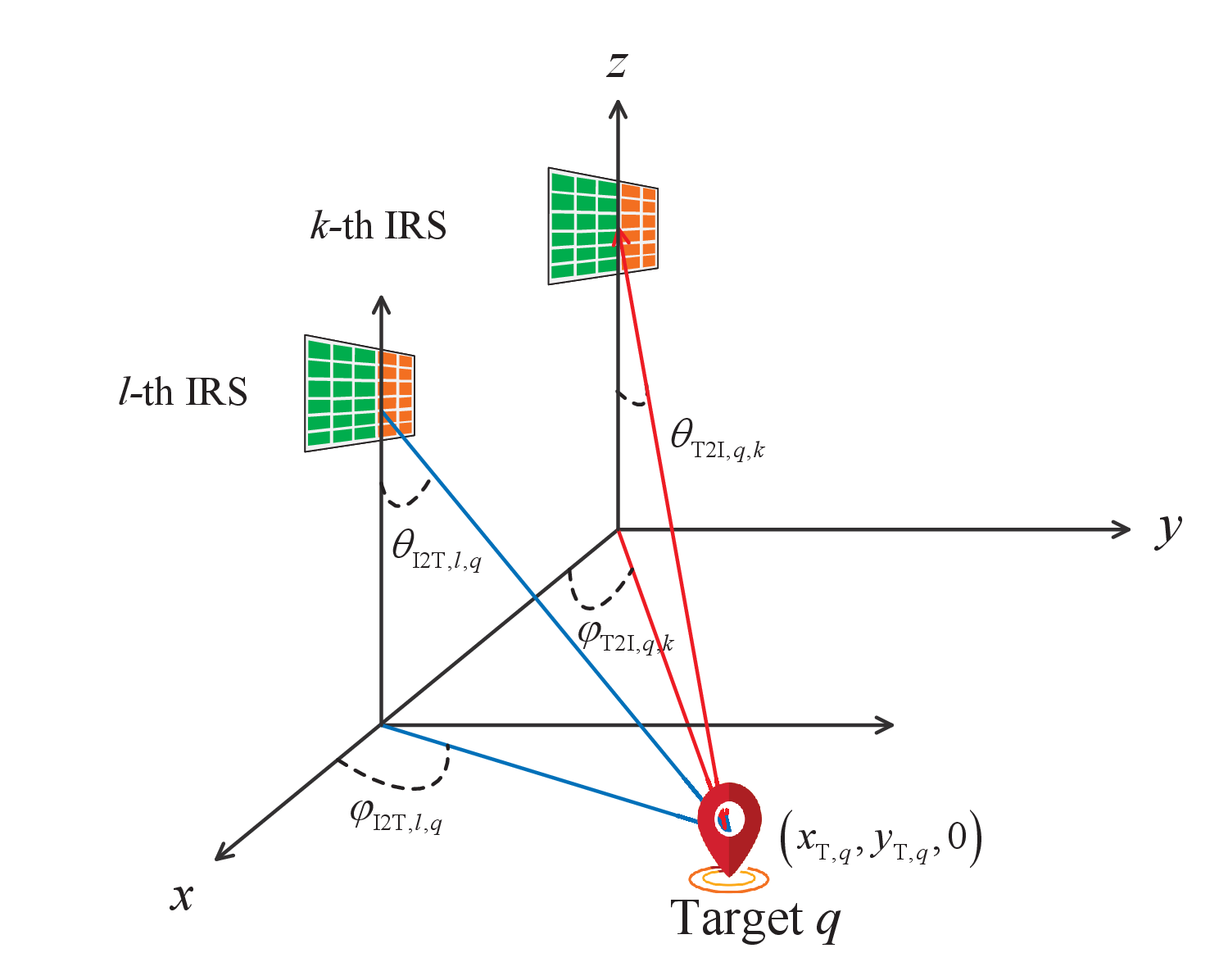}}
	\caption{Relative angle between multiple IRSs and target.}
	\label{angle}
\end{figure}

\begin{figure*}[!t]
	\normalsize
\begin{align}
& \frac{\partial {{\tau }_{q,k,l}}}{\partial {{x}_{\text{T},q}}}=\frac{-1}{c}\left[ \left( \sin {{\varphi }_{\text{I2T},k,q}}\cos {{\theta }_{\text{I2T},k,q}} \right)+\left( \sin {{\varphi }_{\text{T2I},q,l}}\cos {{\theta }_{\text{T2I},q,l}} \right) \right]\triangleq \frac{-1}{c}{{a}_{q,k,l}}, \label{taux}\\ 
& \frac{\partial {{\tau }_{q,k,l}}}{\partial {{y}_{\text{T},q}}}=\frac{-1}{c}\left[ \left( \sin {{\varphi }_{\text{I2T},k,q}}\sin {{\theta }_{\text{I2T},k,q}} \right)+\left( \sin {{\varphi }_{\text{T2I},q,l}}\sin {{\theta }_{\text{T2I},q,l}} \right) \right]\triangleq \frac{-1}{c}{{b}_{q,k,l}}.\label{tauy} 
\end{align}

	\hrulefill
	\vspace*{1pt}
\end{figure*}
Recalling (\ref{fu}), (\ref{fv}), (\ref{vu}) and (\ref{vuq}), the FIM for all $Q$ targets can be represented as

\begin{align}
  & \mathbf{F}\left( \mathbf{u} \right)=\left[ \begin{matrix}
   \frac{\partial \boldsymbol{\tau}}{\partial {{\mathbf{u}}_{1}}}{{\mathbf{F}}_{1,1}}\frac{\partial {{\boldsymbol{\tau}}^{T}}}{\partial {{\mathbf{u}}_{1}}} & \cdots  & {{\mathbf{0}}_{2\times 2}}  \\
   \vdots  & \ddots  & \vdots   \\
   {{\mathbf{0}}_{2\times 2}} & \cdots  & \frac{\partial \boldsymbol{\tau}}{\partial {{\mathbf{u}}_{Q}}}{{\mathbf{F}}_{Q,Q}}\frac{\partial {{\boldsymbol{\tau}}^{T}}}{\partial {{\mathbf{u}}_{Q}}}  \\
\end{matrix} \right] \nonumber\\ 
 & \triangleq \left[ \begin{matrix}
   {{\mathbf{G}}_{1}} & \cdots  & {{\mathbf{0}}_{2\times 2}}  \\
   \vdots  & \ddots  & \vdots   \\
   {{\mathbf{0}}_{2\times 2}} & \cdots  & {{\mathbf{G}}_{Q}}  \\
\end{matrix} \right]. 
\end{align}
For the $q$-th targrt, the CRB of its location $\left( {{x}_{\text{T},q}},{{y}_{\text{T},q}} \right)$ is given by (\ref{crbq}), where $c_0$ is a constant.

\begin{figure*}[!t]
	\normalsize
\begin{align}\label{crbq}
  & \text{CRB}\left( {{x}_{\text{T},q}},{{y}_{\text{T},q}} \right)=\text{tr}\left( \mathbf{G}_{q}^{-1} \right)=c_{0}\text{tr}\left( \left[ \begin{matrix}
   \sum\limits_{k=1}^{K}{\sum\limits_{l=1}^{K}{a_{q,k,l}^{2}\left| {{\mathbf{H}}_{q,k,l}}{{\mathbf{w}}_{k}} \right|_{2}^{2}}} & \sum\limits_{k=1}^{K}{\sum\limits_{l=1}^{K}{{{a}_{q,k,l}}{{b}_{q,k,l}}\left| {{\mathbf{H}}_{q,k,l}}{{\mathbf{w}}_{k}} \right|_{2}^{2}}}  \\
   \sum\limits_{k=1}^{K}{\sum\limits_{l=1}^{K}{{{a}_{q,k,l}}{{b}_{q,k,l}}\left| {{\mathbf{H}}_{q,k,l}}{{\mathbf{w}}_{k}} \right|_{2}^{2}}} & \sum\limits_{k=1}^{K}{\sum\limits_{l=1}^{K}{b_{q,k,l}^{2}\left| {{\mathbf{H}}_{q,k,l}}{{\mathbf{w}}_{k}} \right|_{2}^{2}}}  \\
\end{matrix} \right]^{-1} \right) \nonumber\\ 
 & =c_{0}\frac{\sum\limits_{k=1}^{K}{\sum\limits_{l=1}^{K}{a_{q,k,l}^{2}\left| {{\mathbf{H}}_{q,k,l}}{{\mathbf{w}}_{k}} \right|_{2}^{2}}}+\sum\limits_{k=1}^{K}{\sum\limits_{l=1}^{K}{b_{q,k,l}^{2}\left| {{\mathbf{H}}_{q,k,l}}{{\mathbf{w}}_{k}} \right|_{2}^{2}}}}{\left( \sum\limits_{k=1}^{K}{\sum\limits_{l=1}^{K}{a_{q,k,l}^{2}\left| {{\mathbf{H}}_{q,k,l}}{{\mathbf{w}}_{k}} \right|_{2}^{2}}} \right)\left( \sum\limits_{k=1}^{K}{\sum\limits_{l=1}^{K}{b_{q,k,l}^{2}\left| {{\mathbf{H}}_{q,k,l}}{{\mathbf{w}}_{k}} \right|_{2}^{2}}} \right)-{{\left( \sum\limits_{l=1}^{K}{{{a}_{q,k,l}}{{b}_{q,k,l}}\left| {{\mathbf{H}}_{q,k,l}}{{\mathbf{w}}_{k}} \right|_{2}^{2}} \right)}^{2}}}.
\end{align}
	\hrulefill
	\vspace*{1pt}
\end{figure*}

\subsection{Problem Formulation}
Let $\boldsymbol{\theta}=\left[ {{\boldsymbol{\theta }}_{1}},\cdots ,{{\boldsymbol{\theta }}_{K}} \right]\in {{\mathbb{C}}^{N\times K}}$ and $\mathbf{W}=\left[ {{\mathbf{w}}_{1}},\cdots ,{{\mathbf{w}}_{K}} \right]\in{{\mathbb{C}}^{{{N}_{T}}\times K}}$.
In this paper, we aim to minimize the maximum CRB of estimation of $Q$ targets' location by jointly optimizing the transmit beamforming $\mathbf{W}$
at the BS and reflect beamforming $\mathbf{\theta }$ at the IRSs, subject to maximum transmit power constraints.
The corresponding optimization problem is formulated as
\begin{subequations}\label{p1}
\begin{align}
\text{(P1)}:\quad & \underset{\boldsymbol{\theta} ,\mathbf{W}}{\mathop{\min }}\,\text{ }\underset{q}{\mathop{\text{max}}}\quad\text{CRB}\left( {{x}_{\text{T},q}},{{y}_{\text{T},q}} \right) \\
 \rm{s.t.}\quad &{{\mathbf{H}}_{q,{{k}_{1}},l}}{{\mathbf{w}}_{{k}_{2}}}=\mathbf{0},{{k}_{1}}\ne {{k}_{2}},\forall {{k}_{1}},{{k}_{2}},q,\label{p1a}\\ 
 & \text{tr}\left( \mathbf{W}{{\mathbf{W}}^{H}} \right)={{P}_{\max }},\label{p1b}\\ 
 & 0\le {{\alpha }_{k,n}}\le 1,\forall k, n,\label{p1c}\\ 
 & 0\le {{\beta }_{k,n}}<2\pi ,\forall k, n.\label{p1d}
\end{align}
\end{subequations}
where ${{P}_{\max}}$ denotes the maximum transmitting power of the BS. Note that the objective function is non-convex due to the coupling $\boldsymbol{\theta}$ and $\mathbf{W}$
in targets’ CRB expressions. In addition, the
constraints in (\ref{p1a}) are bilinear equation constraints and can not be tackled by existing methods.
In this paper, we propose a new two-stage algorithm to solve (P1) and show that by applying reasonable reformulations to the CRB expression in (P1), the problem can be efficiently solved with high-quality solutions.

\section{Single-Target System}
In this section, we consider the single-target scenario, i.e., $Q$ = 1, where only one target needs to be localized in the time-frequency domain. In this case, echo signals for localization of different targets
are distinguished in the time or frequency domain. Therefore, (P1) is simplified to 
\begin{subequations}\label{p2}
\begin{align}
\text{(P2)}:\quad & \underset{\boldsymbol{\theta} ,\mathbf{W}}{\mathop{\min }}\quad\text{CRB}\left( {{x}_{\text{T},1}},{{y}_{\text{T},1}} \right) \\
 \rm{s.t.}\quad &{{\mathbf{H}}_{1,{{k}_{1}},l}}{{\mathbf{w}}_{{k}_{2}}}=\mathbf{0},{{k}_{1}}\ne {{k}_{2}},\forall {{k}_{1}},{{k}_{2}},\label{p2a}\\ 
 & \text{tr}\left( \mathbf{W}{{\mathbf{W}}^{H}} \right)={{P}_{\max }},\label{p2b}\\ 
 & 0\le {{\alpha }_{k,n}}\le 1,\forall k, n,\label{p2c}\\ 
 & 0\le {{\beta }_{k,n}}<2\pi ,\forall k, n.\label{p2d}
\end{align}
\end{subequations}
Although much simplified, (P2) is still a non-convex optimization problem because of the (\ref{p2a}) and coupled $\boldsymbol{\theta}$ and $\mathbf{W}$. In this section, we propose a two-stage ADMM-based algorithm
to solve (P2). Specifically, in the first stage,
We activate all IRS reflective elements and achieve 
orthogonality required in (\ref{p2a}) only by active beamforming at BS; in the second stage, we achieve 
orthogonality by joint active and passive beamforming.

\subsection{Reflective Beamforming Design}
First, in the first stage, we do not consider the case 
${{\mathbf{\Theta }}_{k}}={{\mathbf{0}}_{K\times K}}$ and assume ${{\mathbf{a}}^{H}}\left( {{\Omega }_{\text{B2I},{{k}_{1}}}},{{N}_{T}} \right){{\mathbf{w}}_{{k}_{2}}}=0,{{k}_{1}}\ne {{k}_{2}}$. Specifically, (P2) can be reduce to
\begin{subequations}\label{p2.1}
\begin{align}
\text{(P2.1)}:\quad & \underset{\boldsymbol{\theta} ,\mathbf{W}}{\mathop{\min }}\quad\text{CRB}\left( {{x}_{\text{T},1}},{{y}_{\text{T},1}} \right) \\
 \rm{s.t.}\quad &{{\mathbf{a}}^{H}}\left( {{\Omega }_{\text{B2I},{{k}_{1}}}},{{N}_{T}} \right){{\mathbf{w}}_{{k}_{2}}}=0,{{k}_{1}}\ne {{k}_{2}},\\ 
 & \text{(\ref{p2b})}, \text{(\ref{p2c})}, \text{(\ref{p2d})}.\nonumber 
\end{align}
\end{subequations}
Prior to solving (P2.1), we first analyze the relationship between CRB and beam energy.

$\emph{Proposition 1:}$ $\text{CRB}\left( {{x}_{\text{T},1}},{{y}_{\text{T},1}} \right)$ monotonically non-increasing with the increase of $\left\| {{\mathbf{H}}_{1,k,l}}{{\mathbf{w}}_{k}} \right\|_{2}^{2}$.

\emph{Proof}: Let
\begin{align}
{{\mathbf{G}}_{1,k,l}}=\left[ \begin{matrix}
   a_{1,k,l}^{2} & {{a}_{1,k,l}}{{b}_{1,k,l}}  \\
   {{a}_{1,k,l}}{{b}_{1,k,l}} & b_{1,k,l}^{2}  \\
\end{matrix} \right]\in {{\mathbb{R}}^{2\times 2}}.
\end{align}
${{\mathbf{G}}_{1}}$ can be represented as ${{\mathbf{G}}_{1}}=\sum\limits_{k=1}^{K}{\sum\limits_{l=1}^{K}{\left| {{\mathbf{H}}_{1,k,l}}{{\mathbf{w}}_{k}} \right|_{2}^{2}{{\mathbf{G}}_{1,k,l}}}}$. It's obvious that ${{\mathbf{G}}_{1,k,l}}$ is a semi-positive definite matrix and $\left| {{\mathbf{H}}_{1,k,l}}{{\mathbf{w}}_{k}} \right|_{2}^{2}\ge 0$. Therefore, 
${{\mathbf{G}}_{1}}$ is a semi-positive definite matrix with two eigenvalues are ${{\lambda }_{1}}$ and ${{\lambda }_{2}}$, ${{\lambda }_{1}}\ge {{\lambda }_{2}}\ge 0$,${{\lambda }_{1}}{{\lambda }_{2}}\ne 0$.

Without loss of generality, we assume that the increment of $\left\| {{\mathbf{H}}_{1,k,l}}{{\mathbf{w}}_{k}} \right\|_{2}^{2}$ is $\Delta \left| {{\mathbf{H}}_{1,k,l}}{{\mathbf{w}}_{k}} \right|_{2}^{2}$
and the corresponding matrix is $\mathbf{G}_{1}^{'}$ after the increase, which can be represented as
\begin{align}
\mathbf{G}_{1}^{'}={{\mathbf{G}}_{1}}+\Delta \left| {{\mathbf{H}}_{1,k,l}}{{\mathbf{w}}_{k}} \right|_{2}^{2}{{\mathbf{G}}_{1,k,l}}.
\end{align}
Similarly, ${\mathbf{G}_{1}^{'}}$ is a semi-positive definite matrix with two eigenvalues are ${{\lambda }_{1}^{'}}$ and ${{\lambda }_{2}^{'}}$, ${{\lambda }_{1}^{'}}\ge {{\lambda }_{2}^{'}}\ge 0$,$\lambda _{1}^{'}\lambda _{2}^{'}\ne 0$. According to Weyl inequality, we have $\lambda _{1}^{'}\ge {{\lambda }_{1}},\lambda _{2}^{'}\ge {{\lambda }_{2}}$. Therefore,
\begin{align}
\text{tr}\left( {{\left( \mathbf{G}_{\mathbf{1}}^{\mathbf{'}} \right)}^{-1}} \right)=\frac{1}{\lambda _{1}^{'}}+\frac{1}{\lambda _{2}^{'}}\le \frac{1}{{{\lambda }_{1}}}+\frac{1}{{{\lambda }_{2}}}=\text{tr}\left( \mathbf{G}_{1}^{-1} \right).
\end{align}
The proof is completed.$\hfill\blacksquare$

Based on Proposition 1, the design of ${{\mathbf{\Theta }}_{k}}$ is supposed to maximize $\left\| {{\mathbf{H}}_{1,k,l}}{{\mathbf{w}}_{k}} \right\|_{2}^{2}$, which can be represented as
$\alpha _{\text{I2I},1,k,l}^{2}M\left\| {{\mathbf{a}}_{R}^{H}}\left( {{\Phi }_{\text{I2T},k,1}},{{\Omega }_{\text{I2T},k,1}} \right)\text{diag}\left( {{\mathbf{H}}_{\text{B2I},k}}{{\mathbf{w}}_{k}} \right){\boldsymbol{\theta }_{k}} \right\|_{2}^{2}$.
Accordingly, the optimal design of ${{\mathbf{\Theta }}_{k}}$ is given by
\begin{align}\label{thetaopt}
{\boldsymbol{\theta }_{k}}=\frac{{{\left( \mathbf{a}_{R}^{H}\left( {{\Phi }_{\text{I2T},k,1}},{{\Omega }_{\text{I2T},k,1}} \right)\text{diag}\left( {{\mathbf{H}}_{\text{B2I},k}}{{\mathbf{w}}_{k}} \right) \right)}^{*}}}{\left| \mathbf{a}_{R}^{H}\left( {{\Phi }_{\text{I2T},k,1}},{{\Omega }_{\text{I2T},k,1}} \right)\text{diag}\left( {{\mathbf{H}}_{\text{B2I},k}}{{\mathbf{w}}_{k}} \right) \right|}.
\end{align}
\subsection{Transmitting Beamforming Design}
In the first stage, the transmitting beamforming design is supposed to satisfy ${{\mathbf{a}}^{H}}\left( {{\Omega }_{\text{B2I},{{k}_{1}}}},{{N}_{T}} \right){{\mathbf{w}}_{{k}_{2}}}=0,{{k}_{1}}\ne {{k}_{2}}$. Therefore, the feasible 
design is given by
\begin{align}\label{w}
\mathbf{W}=\mathbf{A}_{\text{B2I}}^{H}{{\left( {{\mathbf{A}}_{\text{B2I}}}\mathbf{A}_{\text{B2I}}^{H} \right)}^{-1}}\mathbf{E},
\end{align}
where ${{\mathbf{A}}_{\text{B2I}}}=\left[ \mathbf{a}_{\text{B2I},1}^{H},\cdots ,\mathbf{a}_{\text{B2I},K}^{H} \right]\in {{\mathbb{C}}^{K\times {{N}_{T}}}}$ and $\mathbf{E}=\text{diag}\left( \mathbf{e} \right)=\text{diag}\left( \left[ {{e}_{1}},\cdots ,{{e}_{k}} \right] \right)\in {{\mathbb{R}}^{K\times K}}$ 
denotes the transmitting power allocation over different beams. According to the transmitting power constraint (\ref{p1b}), we have 

Let ${{\alpha }_{k,l}}=\alpha _{\text{I2I},1,k,l}^{2}\alpha _{\text{B2I},k}^{2}M{{N}^{2}}$, ${{a}_{k}}=\sum\limits_{l=1}^{K}{a_{1,k,l}^{2}}{{\alpha }_{k,l}}$, 
${{b}_{k}}=\sum\limits_{l=1}^{K}{b_{1,k,l}^{2}}{{\alpha }_{k,l}}$ and 
${{c}_{k}}=\sum\limits_{l=1}^{K}{{{a}_{1,k,l}}}{{b}_{1,k,l}}{{\alpha }_{k,l}}$. Based on (\ref{thetaopt}), (\ref{w}) and above definition, (P2.1) is equivalent to
\begin{subequations}\label{p22}
\begin{align}
\text{(P2.2)}:\quad & \underset{\mathbf{e}}{\mathop{\min }}\quad \frac{\sum\limits_{k=1}^{K}{e_{k}^{2}{{a}_{k}}}+\sum\limits_{k=1}^{K}{e_{k}^{2}{{b}_{k}}}}{\left( \sum\limits_{k=1}^{K}{e_{k}^{2}{{a}_{k}}} \right)\left( \sum\limits_{k=1}^{K}{e_{k}^{2}{{b}_{k}}} \right)-{{\left( \sum\limits_{k=1}^{K}{e_{k}^{2}{{c}_{k}}} \right)}^{2}}} \\
 \rm{s.t.}\quad & \sum\limits_{k=1}^{K}{e_{k}^{2}}{{\left[ {{\left( {{\mathbf{A}}_{\text{B2I}}}\mathbf{A}_{\text{B2I}}^{H} \right)}^{-1}} \right]}_{k,k}}={{P}_{\max }},
\end{align}
\end{subequations}
which is also difficult to solve because it is a fractional programming and the denominator is non-convex.
For the convenience of representation, we use vector form to equivalently represent the problem as
\begin{subequations}\label{p23}
\begin{align}
\text{(P2.3)}:\quad & \underset{\mathbf{p}}{\mathop{\min }}\quad \frac{\left( {{\mathbf{a}}^{T}}+{{\mathbf{b}}^{T}} \right)\mathbf{p}}{{{\mathbf{p}}^{T}}\left( \mathbf{a}{{\mathbf{b}}^{T}}-\mathbf{c}{{\mathbf{c}}^{T}} \right)\mathbf{p}} \\
 \rm{s.t.}\quad & {{\mathbf{h}}^{T}}\mathbf{p}={{P}_{\max }},\\
 & {{p}_{k}}\ge 0,\forall k,
\end{align}
\end{subequations}
where $\mathbf{p}={{\left[ e_{1}^{2},\cdots e_{K}^{2} \right]}^{T}}$, ${{p}_{k}}$ denotes the $k$-th element in $\mathbf{p}$, $\mathbf{a}={{\left[ {{a}_{1}},\cdots ,{{a}_{K}} \right]}^{T}}$, $\mathbf{b}={{\left[ {{b}_{1}},\cdots ,{{b}_{K}} \right]}^{T}}$, $\mathbf{c}={{\left[ {{c}_{1}},\cdots ,{{c}_{K}} \right]}^{T}}$, $\mathbf{h}={{\left[ {{\left[ {{\left( {{\mathbf{A}}_{\text{B2I}}}\mathbf{A}_{\text{B2I}}^{H} \right)}^{-1}} \right]}_{1,1}},\cdots ,{{\left[ {{\left( {{\mathbf{A}}_{\text{B2I}}}\mathbf{A}_{\text{B2I}}^{H} \right)}^{-1}} \right]}_{K,K}} \right]}^{T}}$. Since $\mathbf{a}{{\mathbf{b}}^{T}}-\mathbf{c}{{\mathbf{c}}^{T}}$ is an indefinite matrix, we introduce additional auxiliary $\mathbf{z}$ to deal with non-convexity in the denominator and penalty term $\frac{\rho }{2} \left\| \mathbf{p}-\mathbf{z} \right\|_{2}^{2}$ to ensure equivalence with the original expression. Specifically, ${{\mathbf{p}}^{T}}\left( \mathbf{a}{{\mathbf{b}}^{T}}-\mathbf{c}{{\mathbf{c}}^{T}} \right)\mathbf{p}$ can be represented as 
\begin{align}\label{de}
{{f}_{d}}\left( \mathbf{p},\mathbf{z} \right)\triangleq {{\mathbf{p}}^{T}}\left( \mathbf{A}-\beta  \right)\mathbf{p}+\beta {{\mathbf{z}}^{T}}\mathbf{z}-\frac{\rho }{2} \left\| \mathbf{p}-\mathbf{z} \right\|_{2}^{2},
\end{align}
where $\mathbf{A}=\frac{1}{2}\left( \mathbf{a}{{\mathbf{b}}^{T}}+\mathbf{b}{{\mathbf{a}}^{T}} \right)$ and $\beta$ is the maximum eigenvalue of $\mathbf{A}$, ${\rho }/{2}$ is a penalty factor and should be greater than $\beta$. The above restriction on the $\beta$ and ${\rho }$ ensures that (\ref{de}) is a concave function with respect to $\mathbf{p}$ and $\mathbf{z}$.
Then, by utilizing the Dinkelbach’s transform, the original fractional planning can be transformed into the following form
\begin{subequations}\label{p24}
\begin{align}
\text{(P2.4)}:\quad & \underset{\mathbf{p},\mathbf{z}}{\mathop{\min }}\quad \alpha \left( {{\mathbf{a}}^{T}}+{{\mathbf{b}}^{T}} \right)\mathbf{p}-{{f}_{d}}\left( \mathbf{p},\mathbf{z} \right)\\
 \rm{s.t.}\quad & {{\mathbf{h}}^{T}}\mathbf{z}={{P}_{\max }},\label{zcon1}\\
 & {{z}_{k}}\ge 0,\forall k,\label{zcon2}\\
 & \mathbf{p}=\mathbf{z}, \label{eqcon}
\end{align}
\end{subequations}
where ${{z}_{k}}$ is the $k$-th element in $\mathbf{z}$
and $\alpha$ is a new auxiliary variable, which is updated by 
\begin{equation}
\label{updatea}
\alpha \left[ t+1 \right]=\frac{{{f}_{d}}\left( \mathbf{p}\left[ t \right],\mathbf{z}\left[ t \right] \right)}{\left( {{\mathbf{a}}^{T}}+{{\mathbf{b}}^{T}} \right)\mathbf{p}\left[ t \right]},
\end{equation}
where $t$ is the iteration index. To solve (P2.4), we employ ADMM method to handle the equation constraints (\ref{eqcon}). The augmented Lagrangian function of (P2.4) can be expressed as
\begin{flalign}
L\left( \mathbf{p},\mathbf{z},\boldsymbol{\lambda } \right)=\alpha \left( {{\mathbf{a}}^{T}}+{{\mathbf{b}}^{T}} \right)\mathbf{p}-{{f}_{d}}\left( \mathbf{p},\mathbf{z} \right)+{{\boldsymbol{\lambda }}^{T}}\left( \mathbf{p}-\mathbf{z} \right), 
\end{flalign}
where ${\boldsymbol{\lambda }}$ is Lagrange multiplier. The variables can be divided into three blocks and updated sequentially by solving the corresponding subproblems. Specifically, the subproblems and their solutions are elaborated as follows.

1) Subproblem with respect to $\mathbf{p}$

For optimizing $\mathbf{p}$, we consider the following subproblem
\begin{align}
\underset{\mathbf{p}}{\mathop{\min }}\quad \alpha \left( {{\mathbf{a}}^{T}}+{{\mathbf{b}}^{T}} \right)\mathbf{p}-{{f}_{d}}\left( \mathbf{p},\mathbf{z} \right)+{{\boldsymbol{\lambda }}^{T}}\left( \mathbf{p}-\mathbf{z} \right),
\end{align}
which is a standard convex problem and can be solved by 
the first order optimality condition, i.e., 
\begin{align}\label{updatep}
& \mathbf{p}\left[ s+1 \right]={{\left( \mathbf{a}{{\mathbf{b}}^{T}}-\mathbf{c}{{\mathbf{c}}^{T}}-\beta \mathbf{I}-\frac{\rho }{2}\mathbf{I} \right)}^{-1}}\nonumber\\ 
& \cdot \left( \frac{\alpha \left( \mathbf{a}+\mathbf{b} \right)+\boldsymbol{\lambda }\left[ s \right]-\rho \mathbf{z}\left[ s \right]}{2} \right) \ 
\end{align}

2) Subproblem with respect to $\mathbf{z}$

For optimizing $\mathbf{z}$ with fixing the other variables, the corresponding subproblem  is reduced to
\begin{subequations}\label{p25}
\begin{align}
\text{(P2.5)}:\quad & \underset{\mathbf{z}}{\mathop{\min }}\quad \frac{\rho }{2}\left\| \mathbf{p}-\mathbf{z} \right\|_{2}^{2}-\beta {{\mathbf{z}}^{T}}\mathbf{z}+{{\boldsymbol{\lambda }}^{T}}\left( \mathbf{p}-\mathbf{z} \right)\\
 \rm{s.t.}\quad &  \text{(\ref{zcon1})},  \text{(\ref{zcon2})} \nonumber
\end{align}
\end{subequations}

The Lagrangian function of this subproblem can be formulated as
\begin{align}
& L\left( \mathbf{z},{{\mu }_{0}},\boldsymbol{\mu } \right)=\frac{\rho }{2}\left\| \mathbf{p}-\mathbf{z} \right\|_{2}^{2}-\beta {{\mathbf{z}}^{T}}\mathbf{z}+{{\boldsymbol{\lambda }}^{T}}\left( \mathbf{p}-\mathbf{z} \right) \nonumber\\ 
& +{{\mu }_{0}}\left( {{\mathbf{h}}^{T}}\mathbf{z}-{{P}_{\max }} \right)+{{\boldsymbol{\mu }}^{T}}\mathbf{z}, 
\end{align}
where ${\mu }_{0}$ and ${\boldsymbol{\mu }}={{\left[ {{\mu}_{1}},\cdots ,{{\mu}_{K}} \right]}^{T}}$
are Lagrange multipliers. (P2.5) is a convex problem and 
can be solved according to Karush-Kuhn-Tucker (KKT) conditions. The optimal solution to (P2.4) is given by
\begin{align}\label{updatez}
{{z}_{i}}=\left\{ \begin{aligned}
  & \frac{\left( {{\lambda }_{i}}+\rho {{p}_{i}}-{{\mu }_{0}}{{h}_{i}} \right)}{2\left( \frac{\rho }{2}-\beta  \right)},{{\mu }_{0}}<\frac{{{\lambda }_{i}}+\rho {{p}_{i}}}{{{h}_{i}}}, \\ 
 & 0,{{\mu }_{0}}\ge \frac{{{\lambda }_{i}}+\rho {{p}_{i}}}{{{h}_{i}}}, \\ 
\end{aligned} \right.
\end{align}
where the value of ${\mu }_{0}$ is determined according to \text{(\ref{zcon1})}. The details of solving this subproblem is shown in Appendix B.

3) Subproblem with respect to ${\boldsymbol{\lambda}}$

The optimization for ${\boldsymbol{\lambda}}$ is based on the multiplier update rule, i.e.,
\begin{align}\label{updatel}
\boldsymbol{\lambda }\left[ s+1 \right]=\boldsymbol{\lambda }\left[ s \right]+\rho \left( \mathbf{p}\left[ s+1 \right]-\mathbf{z}\left[ s+1 \right] \right).
\end{align}

\subsection{Two-Stage Algorithm}\label{selecetion}
So far, we have solved all subproblems with closed-form
solutions and the proposed ADMM-based algorithm for
(P2.1) is summarized in \textbf{Algorithm 1}.

\begin{algorithm}[t]
	\caption{Proposed ADMM-based Algorithm.} %算法的名字
	\begin{algorithmic}[1]
		\State$\textbf{Input}$: Initialize $\alpha$,  $\mathbf{p}$, $\mathbf{z}$,  $\boldsymbol{\lambda }$, $\rho$, $\beta $, threshold ${{\varepsilon }_{1}}$, ${{\varepsilon }_{2}}$ and iteration index $t=1$ for outer loop
		\Repeat: outer loop
        \State Initialize iteration index $s=1$ for inner loop.
		\Repeat: inner loop
        \State Update $\mathbf{p}$ according to (\ref{updatep}). 
        \State Update $\mathbf{z}$ according to (\ref{updatez}). 
        \State Update $\boldsymbol{\lambda }$ according to (\ref{updatel}). 
        \State $s=s+1$. 
		\Until 
$\left\| \mathbf{p}\left[ s \right]-\mathbf{p}\left[ s-1 \right] \right\|_{2}^{2}\le {{\varepsilon }_{1}}$,
$\left\| \mathbf{z}\left[ s \right]-\mathbf{z}\left[ s-1 \right] \right\|_{2}^{2}\le {{\varepsilon }_{1}}$,
$\left\| \mathbf{p}\left[ s \right]-\mathbf{z}\left[ s \right] \right\|_{2}^{2}\le {{\varepsilon }_{1}}$.
        \State $t=t+1$.
        \State Updata $\alpha$ according to (\ref{updatea}). 
        \Until $\alpha \left[ t \right]-\alpha \left[ t-1 \right]\le {{\varepsilon }_{2}}$.
	\end{algorithmic}
\end{algorithm}

In the first stage, we assume ${{\mathbf{a}}^{H}}\left( {{\Omega }_{\text{B2I},{{k}_{1}}}},{{N}_{T}} \right){{\mathbf{w}}_{{k}_{2}}}=0,{{k}_{1}}\ne {{k}_{2}}$ to satisfy (\ref{p2a}), which means orthogonality is achieved only by transmitting beamforming. In fact, we find reflective elements in some IRSs may not be are assigned power after solving (P2.1), i.e., $\exists \tilde{k},e_{{\tilde{k}}}^{2}=0$. These activated reflective elements do not benefit the system performance but increase the difficulty of achieving channel orthogonality.

Therefore, in the second stage, we can relax the channel orthogonality constraint by not activating reflective elements on some IRSs. 
Specifically, after solving (P2.1), we divide $\mathcal{K}$ into $\widetilde{\mathcal{K}}=\left\{ \tilde{k}|e_{{\tilde{k}}}^{2}=0,\tilde{k}\in \mathcal{K} \right\}$ and $\overline{\mathcal{K}}=\left\{ \bar{k}|e_{{\bar{k}}}^{2}>0,\bar{k}\in \mathcal{K} \right\}$,$\tilde{\mathcal{K}}+\bar{\mathcal{K}}=\mathcal{K}$. Then, we set ${{\mathbf{\Theta }}_{{\tilde{k}}}}={{\mathbf{0}}_{N\times N}},{{\mathbf{w}}_{{\tilde{k}}}}={{\mathbf{0}}_{{{N}_{T}}\times 1}},\forall \tilde{k}\in \widetilde{\mathcal{K}}$ and resolve (P2.1) according to \textbf{Algorithm 1}  while reducing the number of variables to be optimized (change subscript set from $\forall k\in \mathcal{K}$ to $\forall\bar{k}\in \bar{\mathcal{K}}$). We repeat the above process until one time after solving (P2.1), all activated reflective elements are assigned power.

\textbf{Algorithm 1} is computationally efficient because all variables are updated via closed-form expressions. Specifically, the complexity of updating 
$\mathbf{p}$, $\mathbf{z}$ and ${\boldsymbol{\lambda}}$ is ${\mathcal{O}}\left( {{K}^{3}} \right)$,
${\mathcal{O}}\left( {{I}_{in}}K \right)$ and ${\mathcal{O}}\left( K \right)$, respectively where ${{I}_{in}}$ is the number of inner iterations required for convergence. Therefore, the overall complexity of \textbf{Algorithm 1} is ${\mathcal{O}}\left( {{I}_{out}}\left( {{K}^{3}}+{{I}_{in}}K \right) \right)$
where ${{I}_{out}}$ is the number of outer iterations required for convergence.

\section{Multi-Target System}
In this section, we study the general multi-target localization scenario where multiple targets are located at arbitrary locations and BS and IRSs utilize space beamforming to locate multiple targets in the same time-frequency dimension. For this general scenario, we propose an algorithm based on alternating optimization to solve the (P1).
\subsection{Dual Problem}
By introducing an auxiliary $\gamma$ variable to
characterize the worst-case CRB among multi-target, (P1) can be equivalently rewritten as
\begin{subequations}\label{p31}
\begin{align}
\text{(P3.1)}:\quad & \underset{\boldsymbol{\theta} ,\mathbf{W}}{\mathop{\min }}\quad \frac{1}{\gamma } \\
 \rm{s.t.}\quad &\text{CRB}\left( {{x}_{\text{T},q}},{{y}_{\text{T},q}} \right)\le \frac{1}{\gamma },\forall q,\\ 
 & \text{(\ref{p1a})},\text{(\ref{p1b})},\text{(\ref{p1c})},\text{(\ref{p1d})}.\nonumber
\end{align}
\end{subequations}
The QoS problem and max-min fairness (MMF) problem in communication are a pair of dual problems \cite{7874154}. Similarly, we apply the above conclusions to the design of multi-target localization systems and formulate the following problem.
\begin{subequations}\label{p32}
\begin{align}
\text{(P3.2)}:\quad & \underset{\boldsymbol{\theta} ,\mathbf{W}}{\mathop{\min }}\quad \text{tr}\left( \mathbf{W}{{\mathbf{W}}^{H}} \right) \\
 \rm{s.t.}\quad &\text{CRB}\left( {{x}_{\text{T},q}},{{y}_{\text{T},q}} \right)\le \frac{1}{\varphi },\forall q,\label{p32a}\\ 
& \text{(\ref{p1a})},\text{(\ref{p1c})},\text{(\ref{p1d})}.\nonumber
\end{align}
\end{subequations}
The duality between (P3.1) and (P3.2) can be described as
\begin{align}
\text{P}3.1\left( {{P}_{\max }}={{P}_{0}} \right)=\frac{1}{{{\gamma }_{0}}},\text{P}3.2\left( \frac{1}{\varphi }=\frac{1}{{{\gamma }_{0}}} \right)={{P}_{0}}.
\end{align}
This duality implies that if we can obtain the solution for (P3.2), we can find the solution for (P3.1) by iteratively solving (P3.2) using a bisection search method until the transmit power equals ${{P}_{\max }}$. Moreover, 
$\boldsymbol{\theta}$ and $\mathbf{W}$ are coupled in \text{(\ref{p1a}}) and \text{(\ref{p32a}}). Therefore, we utilize the two-stage algorithm to simplify (P3.2) similar to the single-target system and alternating optimization to solve it. 

\subsection{Transmitting Beamforming Design}
To satisfy orthogonality constraint ${{\mathbf{a}}^{H}}\left( {{\Omega }_{\text{B2I},{{k}_{1}}}},{{N}_{T}} \right){{\mathbf{w}}_{{k}_{2}}}=0,{{k}_{1}}\ne {{k}_{2}}$, we still adopt the form of feasible solutions given in (\ref{w}). Therefore, $\left\| {{\mathbf{H}}_{q,k,l}}{{\mathbf{w}}_{k}} \right\|_{2}^{2}$ can be represented as $\alpha _{\text{I2I},q,k,l}^{2}\alpha _{\text{B2I},k}^{2}e_{k}^{2}M\left\| \mathbf{a}_{\text{I2T},k,q}^{H}{{\mathbf{\Theta }}_{k}}{{\mathbf{a}}_{\text{I2B},k}} \right\|_{2}^{2}$.

Let ${{\alpha }_{q,k,l}}=\alpha _{\text{I2I},q,k,l}^{2}\alpha _{\text{B2I},k}^{2}M$, ${{a}_{q,k}}=\sum\limits_{l=1}^{K}{a_{q,k,l}^{2}}{{\alpha }_{q,k,l}}$, 
${{b}_{q,k}}=\sum\limits_{l=1}^{K}{b_{q,k,l}^{2}}{{\alpha }_{q,k,l}}$ and 
${{c}_{q,k}}=\sum\limits_{l=1}^{K}{{{a}_{q,k,l}}}{{b}_{q,k,l}}{{\alpha }_{q,k,l}}$. By applying
the change of variables $\left\| \mathbf{a}_{\text{I2T},k,q}^{H}{{\mathbf{\Theta }}_{k}}{{\mathbf{a}}_{\text{I2B},k}} \right\|_{2}^{2}=\left\| {{\boldsymbol{\theta }}_{k}^{T}}\text{diag}\left( \mathbf{a}_{\text{I2T},k,q}^{H} \right){{\mathbf{a}}_{\text{I2B},k}} \right\|_{2}^{2}\triangleq {{x}_{q,k}}$, problem (P3.2) is reduced to
\begin{subequations}\label{p33}
\begin{align}
\text{(P3.3)}:\quad & \underset{\mathbf{p}}{\mathop{\min }}\quad{{\mathbf{h}}^{T}}\mathbf{p}\\ 
 \rm{s.t.}\quad & \varphi  \left( \mathbf{a}_{q}^{T}\mathbf{p}+\mathbf{b}_{q}^{T}\mathbf{p} \right)-\mathbf{a}_{q}^{T}\mathbf{pb}_{q}^{T}\mathbf{p}+\mathbf{c}_{q}^{T}\mathbf{pc}_{q}^{T}\mathbf{p}\le 0,\forall q,\label{p33a}\\ 
 & {{p}_{k}}\ge 0,\forall k, 
\end{align}
\end{subequations}
where $\mathbf{p}={{\left[ e_{1}^{2},\cdots e_{K}^{2} \right]}^{T}}$, ${{p}_{k}}$ denotes the $k$-th element in $\mathbf{p}$, ${{\mathbf{a}}_{q}}={{\left[ {{a}_{q,1}}{{x}_{q,1}},\cdots ,{{a}_{q,K}{{x}_{q,k}}} \right]}^{T}}$,
${{\mathbf{b}}_{q}}={{\left[ {{b}_{q,1}}{{x}_{q,1}},\cdots ,{{b}_{q,K}}{{x}_{q,k}} \right]}^{T}}$ and
${{\mathbf{c}}_{q}}={{\left[ {{c}_{q,1}}{{x}_{q,1}},\cdots ,{{x}_{q,k}}{{x}_{q,k}} \right]}^{T}}$.
Note that (P3.3) is an indefinite quadratic optimization problem because of constrain (\ref{p33a}). Therefore, we adopt successive
convex approximation (SCA) technique to approximate the indefinite quadratic term in constrain
(\ref{p33a}). Specifically, in the $r$-th iteration, for a given feasible point ${{\mathbf{p}}_{\left(r \right)}}$, we have
\begin{align}
  & \mathbf{a}_{q}^{T}\mathbf{pb}_{q}^{T}\mathbf{p}-\mathbf{c}_{q}^{T}\mathbf{pc}_{q}^{T}\mathbf{p}={{\mathbf{p}}^{T}}\left( {{\mathbf{A}}_{q}}-{{\beta }_{q}}\mathbf{I} \right)\mathbf{p}+{{\beta }_{q}}{{\mathbf{p}}^{T}}\mathbf{p} \nonumber\\ 
 & \ge {{\mathbf{p}}^{T}}\left( {{\mathbf{A}}_{q}}-{{\beta }_{q}}\mathbf{I} \right)\mathbf{p}+{{\beta }_{q}}\mathbf{p}_{\left( r \right)}^{T}{{\mathbf{p}}_{\left( r \right)}}+2{{\beta }_{q}}\mathbf{p}_{\left( r \right)}^{T}\left( \mathbf{p}-{{\mathbf{p}}_{\left( r \right)}} \right),
\end{align}
where ${{\mathbf{A}}_{q}}=\frac{1}{2}\left( {{\mathbf{a}}_{q}}\mathbf{b}_{q}^{T}+{{\mathbf{b}}_{q}}\mathbf{a}_{q}^{T} \right)$ and
${{\beta }_{q}}$ is the maximum eigenvalue of ${{\mathbf{A}}_{q}}$. 
By replacing the corresponding terms in the
constrains (\ref{p33a}), (P3.3) can be transformed to a convex problem and solved 
by convex optimization tool such as CVX.

\subsection{Reflective Beamforming Design}
For given transmitting beamforming $\mathbf{W}$,
problem (P3.2) is reduced to a feasibility-check problem as follows.
\begin{subequations}\label{p34}
\begin{align}
\text{(P3.4)}:\quad &{\text{Find }}\quad \boldsymbol{\theta}_k \\
 \rm{s.t.}\quad  & \varphi \left( \sum\limits_{k=1}^{K}{{{a}_{q,k}}{{p}_{k}}{{x}_{q,k}}}+\sum\limits_{k=1}^{K}{{{b}_{q,k}}{{p}_{k}}{{x}_{q,k}}} \right) \nonumber\\ 
 & -\sum\limits_{k=1}^{K}{{{a}_{q,k}}{{p}_{k}}{{x}_{q,k}}}\sum\limits_{k=1}^{K}{{{b}_{q,k}}{{p}_{k}}{{x}_{q,k}}}\nonumber \\ 
 & +\sum\limits_{k=1}^{K}{{{c}_{q,k}}{{p}_{k}}{{x}_{q,k}}}\sum\limits_{k=1}^{K}{{{c}_{q,k}}{{p}_{k}}{{x}_{q,k}}}\le 0,\forall q, \label{p34a}\\ 
&\text{(\ref{p1c})},\text{(\ref{p1d})},\nonumber
\end{align}
\end{subequations}
where ${{x}_{q,k}}=\left\| {{\boldsymbol{\theta }}_{k}}\text{diag}\left( \mathbf{a}_{\text{I2T},k,q}^{H} \right){{\mathbf{a}}_{\text{I2B},k}} \right\|_{2}^{2}$. Note that ${{\boldsymbol{\theta }}_{k}}$ appears in quartic form in (\ref{p34a}) and has non-convex
unit-modulus constraints. Therefore, we apply SDR and SCA technique to relax (\ref{p34a}) and solve (P3.4) with a high-quality approximate solution. Specifically, by defining ${{\mathbf{R}}_{k}}={{\boldsymbol{\theta }}_{k}}\boldsymbol{\theta }_{k}^{H}$ and 
${{\mathbf{V}}_{q,k}}=\text{diag}\left( \mathbf{a}_{\text{I2T},k,q}^{H} \right){{\mathbf{a}}_{\text{I2B},k}}\mathbf{a}_{\text{I2B},k}^{H}\text{diag}\left( {{\mathbf{a}}_{\text{I2T},k,q}} \right)$, 
${{x}_{q,k}}$ is equivalent to $\text{tr}\left( {{\mathbf{R}}_{k}}{{\mathbf{V}}_{q,k}} \right)$ with constrains ${{\mathbf{R}}_{k}}\succeq 0,\text{rank}\left( {{\mathbf{R}}_{k}} \right)=1$.
Based on the above equivalent transformation,
we have
\begin{align}
{{x}_{q,k}}{{x}_{q,k'}}=\text{tr}\left( {{\mathbf{R}}_{k}}{{\mathbf{V}}_{q,k}} \right)\text{tr}\left( {{\mathbf{R}}_{k'}}{{\mathbf{V}}_{q,k'}} \right),\forall k,k'\in{\mathcal{K}},
\end{align}
which is a non-convex function because of its
bilinear form with respect to ${{\mathbf{R}}_{k}}$ and ${{\mathbf{R}}_{k'}}$.
Therefore, we replace $\text{tr}\left( {{\mathbf{R}}_{k}}{{\mathbf{V}}_{q,k}} \right)\text{tr}\left( {{\mathbf{R}}_{k'}}{{\mathbf{V}}_{q,k'}} \right)$ equivalently by $f\left( {{\mathbf{R}}_{k}},{{\mathbf{R}}_{k'}} \right)-g\left( {{\mathbf{R}}_{k}},{{\mathbf{R}}_{k'}} \right)$ with 
\begin{align}
  & f\left( {{\mathbf{R}}_{k}},{{\mathbf{R}}_{{{k}'}}} \right)={{\left( \text{tr}\left( {{\mathbf{R}}_{k}}{{\mathbf{V}}_{q,k}} \right)+\left( {{\mathbf{R}}_{{{k}'}}}{{\mathbf{V}}_{q,{k}'}} \right) \right)}^{2}}/4,  \\ 
 & g\left( {{\mathbf{R}}_{k}},{{\mathbf{R}}_{{{k}'}}} \right)={{\left( \text{tr}\left( {{\mathbf{R}}_{k}}{{\mathbf{V}}_{q,k}} \right)-\text{tr}\left( {{\mathbf{R}}_{{{k}'}}}{{\mathbf{V}}_{q,{k}'}} \right) \right)}^{2}}/4.
\end{align}
\begin{figure*}[!t]
	\normalsize
\begin{align}
  & f\left( {{\mathbf{R}}_{k}},{{\mathbf{R}}_{k'}} \right)\ge f\left( \mathbf{R}_{k}^{\left( r \right)},\mathbf{R}_{k}^{\left( r \right)} \right)+2\text{tr}\left( \mathbf{R}_{k}^{\left( r \right)}{{\mathbf{V}}_{q,k}}+\mathbf{R}_{k'}^{\left( r \right)}{{\mathbf{V}}_{q,k'}} \right)\text{tr}\left( {{\mathbf{V}}_{q,k}}\left( {{\mathbf{R}}_{k}}-\mathbf{R}_{k}^{\left( r \right)} \right)+{{\mathbf{V}}_{q,k'}}\left( {{\mathbf{R}}_{k'}}-\mathbf{R}_{k'}^{\left( r \right)} \right) \right)\triangleq {{f}_{lb}}\left( {{\mathbf{R}}_{k}},{{\mathbf{R}}_{k'}} \right) ,\label{flb}\\ 
 & g\left( {{\mathbf{R}}_{k}},{{\mathbf{R}}_{k'}} \right)\ge g\left( \mathbf{R}_{k}^{\left( r \right)},\mathbf{R}_{k}^{\left( r \right)} \right)+2\text{tr}\left( \mathbf{R}_{k}^{\left( r \right)}{{\mathbf{V}}_{q,k}}-\mathbf{R}_{k'}^{\left( r \right)}{{\mathbf{V}}_{q,k'}} \right)\text{tr}\left( {{\mathbf{V}}_{q,k}}\left( {{\mathbf{R}}_{k}}-\mathbf{R}_{k}^{\left( r \right)} \right)-{{\mathbf{V}}_{q,k'}}\left( {{\mathbf{R}}_{k'}}-\mathbf{R}_{k'}^{\left( r \right)} \right) \right)\triangleq {{g}_{lb}}\left( {{\mathbf{R}}_{k}},{{\mathbf{R}}_{k'}} \right).\label{glb}
\end{align}
	\hrulefill
	\vspace*{1pt}
\end{figure*}
Next, we apply SCA to linearize $f\left( {{\mathbf{R}}_{k}},{{\mathbf{R}}_{k'}} \right)$ and $g\left( {{\mathbf{R}}_{k}},{{\mathbf{R}}_{k'}} \right)$, which can relax  (\ref{p34a}) to a convex constraint. Specifically, in the $r$-th iteration, for a given feasible point $\left( \mathbf{R}_{k}^{\left( r \right)},\mathbf{R}_{k'}^{\left( r \right)} \right)$, the lower bounds of $f\left( {{\mathbf{R}}_{k}},{{\mathbf{R}}_{k'}} \right)$ and $g\left( {{\mathbf{R}}_{k}},{{\mathbf{R}}_{k'}} \right)$ are given in (\ref{flb}) and (\ref{glb}) by first-order Taylor expansion, respectively. Based on above 
relaxation, the bilinear form in (\ref{p34a})
can be approximated as
% {\setlength\abovedisplayskip{0.2cm}
% \setlength\belowdisplayskip{0.2cm}
\begin{align}
  & {{x}_{q,k}}{{x}_{q,k'}}\le f\left( {{\mathbf{R}}_{k}},{{\mathbf{R}}_{k'}} \right)-{{g}_{lb}}\left( {{\mathbf{R}}_{k}},{{\mathbf{R}}_{k'}} \right), \label{pxx}\\ 
 & -{{x}_{q,k}}{{x}_{q,k'}}\le -{{f}_{lb}}\left( {{\mathbf{R}}_{k}},{{\mathbf{R}}_{k'}} \right)+g\left( {{\mathbf{R}}_{k}},{{\mathbf{R}}_{k'}} \right).\label{nxx} 
\end{align}
By replacing the corresponding terms ${{x}_{q,k}}{{x}_{q,k'}}$ and $-{{x}_{q,k}}{{x}_{q,k'}}$ in
(\ref{p34a}) according to (\ref{pxx}) and (\ref{nxx}), respectively, and ignoring the rank-one constrain, (P3.4) can be reduced to
% {\setlength\abovedisplayskip{0.2cm}
% \setlength\belowdisplayskip{0.2cm}
\begin{subequations}\label{p35}
\begin{align}
\text{(P3.5)}:\enspace &{\text{Find }}\quad {{\mathbf{R}}_{k}} \\
 \rm{s.t.}\enspace    & \varphi \left( \sum\limits_{k=1}^{K}{{{a}_{q,k}}{{p}_{k}}\text{tr}\left( {{\mathbf{R}}_{k}}{{\mathbf{V}}_{q,k}} \right)}+\sum\limits_{k=1}^{K}{{{b}_{q,k}}{{p}_{k}}\text{tr}\left( {{\mathbf{R}}_{k}}{{\mathbf{V}}_{q,k}} \right)} \right) \nonumber\\ 
 & +\sum\limits_{k=1}^{K}{\sum\limits_{k'=1}^{K}{{{a}_{q,k'}}{{b}_{q,k}}{{p}_{k}}{{p}_{k'}}\left[{{g}}\left( {{\mathbf{R}}_{k}},{{\mathbf{R}}_{k'}} \right)-{{f}_{lb}}\left( {{\mathbf{R}}_{k}},{{\mathbf{R}}_{k'}} \right) \right]}} \nonumber\\ 
 & +\sum\limits_{k=1}^{K}{\sum\limits_{k'=1}^{K}{{{p}_{k}}{{p}_{k'}}{{f}_{\text{sgn}}}\left( {{c}_{q,k}},{{c}_{q,{k}'}},{{\mathbf{R}}_{k}},{{\mathbf{R}}_{k'}} \right)}}\le 0,\forall q, \label{p35a} \\
& {{\left[ {{\mathbf{R}}_{k}} \right]}_{n,n}}=1,\forall k,\forall n,\label{p35b}\\
& {{\mathbf{R}}_{k}}\succeq 0,\forall k,\label{p35c}
\end{align}
\end{subequations}
where ${{f}_{\text{sgn}}}({{c}_{q,k}},{{c}_{q,{k}'}},{{\mathbf{R}}_{k}},{{\mathbf{R}}_{k'}})$ is defined as
% {\setlength\abovedisplayskip{0.2cm}
% \setlength\belowdisplayskip{0.2cm}
\begin{align}
  & {{f}_{\text{sgn}}}({{c}_{q,k}},{{c}_{q,{k}'}},{{\mathbf{R}}_{k}},{{\mathbf{R}}_{k'}})= \nonumber\\ 
 & \left\{ \begin{aligned}
  & {{c}_{q,k}},{{c}_{q,{k}'}}\left[ f\left( {{\mathbf{R}}_{k}},{{\mathbf{R}}_{k'}} \right)-{{g}_{lb}}\left( {{\mathbf{R}}_{k}},{{\mathbf{R}}_{{{k}'}}} \right) \right],{{c}_{q,k}}{{c}_{q,{k}'}}\ge 0 \nonumber,\\ 
 & {{c}_{q,k}},{{c}_{q,{k}'}}\left[ {{f}_{lb}}\left( {{\mathbf{R}}_{k}},{{\mathbf{R}}_{{{k}'}}} \right)-g\left( {{\mathbf{R}}_{k}},{{\mathbf{R}}_{{{k}'}}} \right) \right],{{c}_{q,k}}{{c}_{q,{k}'}}<0.\\ 
\end{aligned} \right.\\
\end{align}
(P3.5) is a standard semi-definite programming problem and can be solved by a convex optimization tool such as CVX. Because we ignore the rank-one constrain and may obtain optimal solutions $\mathbf{R}_{k}^{*}$ with $\text{rank}\left( \mathbf{R}_{k}^{*} \right)>1$,  Gaussian randomization can be used to achieve near-optimal performance.

To achieve better convergence, we introduce ``residual variable" ${{r}_{q}}$ to characterize the difference between $\text{CRB}\left( {{x}_{\text{T},q}},{{y}_{\text{T},q}} \right)$ and $\frac{1}{\varphi }$ and transform the  optimization objectives into an explicit function with respect to ${{r}_{q}}$. The modified problem can be expressed as
\begin{subequations}\label{p36}
\begin{align}
\text{(P3.6)}:\enspace &\underset{{{\mathbf{R}}_{k}},\left\{ {{r}_{q}} \right\}}{\mathop{\max }}\quad \sum\limits_{q=1}^{Q}{{{r}_{q}}}\\
 \rm{s.t.}\enspace    & \varphi \left( \sum\limits_{k=1}^{K}{{{a}_{q,k}}{{p}_{k}}\text{tr}\left( {{\mathbf{R}}_{k}}{{\mathbf{V}}_{q,k}} \right)}+\sum\limits_{k=1}^{K}{{{b}_{q,k}}{{p}_{k}}\text{tr}\left( {{\mathbf{R}}_{k}}{{\mathbf{V}}_{q,k}} \right)} \right) \nonumber\\ 
 & +\sum\limits_{k=1}^{K}{\sum\limits_{k'=1}^{K}{{{a}_{q,k'}}{{b}_{q,k}}{{p}_{k}}{{p}_{k'}}\left[g\left( {{\mathbf{R}}_{k}},{{\mathbf{R}}_{k'}} \right)-{{f}_{lb}}\left( {{\mathbf{R}}_{k}},{{\mathbf{R}}_{k'}} \right) \right]}} \nonumber\\ 
 & +\sum\limits_{k=1}^{K}{\sum\limits_{k'=1}^{K}{{{p}_{k}}{{p}_{k'}}{{f}_{\text{sgn}}}\left( {{c}_{q,k}},{{c}_{q,{k}'}},{{\mathbf{R}}_{k}},{{\mathbf{R}}_{k'}} \right)}}+{{r}_{q}}\le 0,\forall q,  \\
& \text{(\ref{p35a})}, \text{(\ref{p35b})}.\nonumber
\end{align}
\end{subequations}
(P3.5) and (P3.6) have the same feasible solution while 
 Solving (P3.6) can make the alternating optimization algorithm converge faster.
 
\subsection{Two-Stage Algorithm}
\begin{algorithm}[t]
	\caption{Proposed Alternating Optimization Algorithm.} %算法的名字
	\begin{algorithmic}[1]
		\State$\textbf{Input}$: Initialize ${{\varphi }_{\min }}$,  ${{\varphi }_{\max }}$, $\varphi =\sqrt{{{\varphi }_{\max }}{{\varphi }_{\min }}}$, threshold ${{\varepsilon }_{1}}$, ${{\varepsilon }_{2}}$ and iteration index $t=1$ for outer loop
		\Repeat: outer loop
        \State Initialize iteration index $s=1$ for inner loop.
		\Repeat: inner loop
        \State Update $\mathbf{p}$ by solving (P3.3). 
        \State Update ${{\mathbf{R}}_{k}}$ by solving (P3.6) . 
        \State $s=s+1$. 
		\Until 
$\frac{\text{tr}\left( \mathbf{W}\left[ t,s-1 \right]\mathbf{W}{{\left[ t,s-1 \right]}^{H}}-\mathbf{W}\left[ t,s \right]\mathbf{W}{{\left[ t,s \right]}^{H}} \right)}{\text{tr}\left( \mathbf{W}\left[ t,s \right]\mathbf{W}{{\left[ t,s \right]}^{H}} \right)}\le {{\varepsilon }_{1}}$.
        \State $t=t+1$.
\If{$\text{tr}\left( \mathbf{W}\left[ t,s \right]\mathbf{W}{{\left[ t,s \right]}^{H}} \right)\le {{P}_{\max }}$}          
        \State ${{\varphi }_{\max }} = \varphi$.
\Else
        \State ${{\varphi }_{\min }} = \varphi$.
\EndIf

\State $\varphi =\sqrt{{{\varphi }_{\max }}{{\varphi }_{\min }}}$.

\Until 
$\frac{\left| \text{tr}\left( \mathbf{W}\left[ t,s \right]\mathbf{W}{{\left[ t,s \right]}^{H}} \right)-{{P}_{\max }} \right|}{{{P}_{\max }}}\ \le {{\varepsilon }_{2}}$.

	\end{algorithmic}
\end{algorithm}

So far, we have solved two subproblems in alternating optimization and the proposed alternating algorithm for problem (P1) is summarized in \textbf{Algorithm 2}, where
$\mathbf{W}\left[ t,s \right]$ is the output of $\mathbf{W}$ in the $t$-th outer loop and the $s$-th loop of \textbf{Algorithm 2}. 

Like the single-target system, we apply a two-stage algorithm framework to the multi-target system and the specific process is similar to section \ref{selecetion} and omit it here.

For \textbf{Algorithm 2}, the complexity is mainly dominated by solving the quadratic constrained quadratic programming and semidefinite programming problem. Specifically, the complexity of solving (P3.3) and (P3.6) is ${\mathcal{O}}\left( {{I}_{in1}}K{{\left( K+1 \right)}^{2}} \right)$ and ${\mathcal{O}}\left( {{I}_{in2}}\left(K{{N}^{2}}Q+{{K}^{2}}{{N}^{4}}{{Q}^{2}}\right) \right)$, respectively where ${{I}_{in1}}$ and ${{I}_{in2}}$ are the number of inner iterations required for convergence. Therefore, the overall complexity of \textbf{Algorithm 2} is $\mathcal{O}\left( {{I}_{out}}\left( {{I}_{in1}}K{{\left( K+1 \right)}^{2}}+{{I}_{in2}}\left(K{{N}^{2}}Q+{{K}^{2}}{{N}^{4}}{{Q}^{2}}\right) \right) \right)$ is the number of outer iterations required for convergence where ${{I}_{out}}$ is the number of inner iterations required for convergence.

\section{Simulation Results}
In this section, we verify the effectiveness of our proposed algorithm multi-IRS collaborative localization system through numerical results. We consider a three-dimensional coordinate system in this section. The BS is located at (0, 0, 50m) while $K$ IRSs and $Q$ targets are located randomly distributed in the rectangular area whose coordinate is $\left( -100\le x\le 100,-100\le y\le100\right)$. Other main system parameters are listed in Table I. We compare the proposed two-stage algorithm with the other three benchmarks:
\begin{itemize}
\item \textbf{One-stage:} All reflective elements on multiple IRSs
are activated and \textbf{Algorithm 1} is applied only once. 

\item \textbf{Equal power allocation:} Power allocation is not designed and all IRSs receive equal energy with optimal phases.

\item \textbf{Random phase shift:} The phase shift design of all reflective elements on the multiple IRSs is randomized with optimal power allocation.
\end{itemize}
\begin{table}[t]
\centering
\small
\caption{Simulation Parameters}
\scalebox{0.88}{
\begin{tabular}{|c|c|c|c|c|c|}
\hline Parameter & Value & Parameter & Value & Parameter & Value\\
\hline${{N}_{t}}$ & $12$ & ${N}$ & $10$ & $M$ & $10$\\
\hline$K$ & $6$ & ${{\text{H}}_{\text{BS}}}/{{\text{H}}_{\text{IRS}}}$ & $50\text{ m}/30\text{ m}$ & $\kappa $ & $7 \text{ dBsm}$\\
\hline${{P}_{\max}}$ & $20\text{ dBW}$ & $\sigma_{l}^{2}$ & $-110\text{ dBm}$ & $\lambda$ & $0.3\text{m}$\\
\hline
\end{tabular}}
\end{table}
\subsection{Single-Target System}
\subsubsection{CRB Versus Number of IRSs}
\begin{figure}%[htbp]
\centerline{\includegraphics[width=8cm]{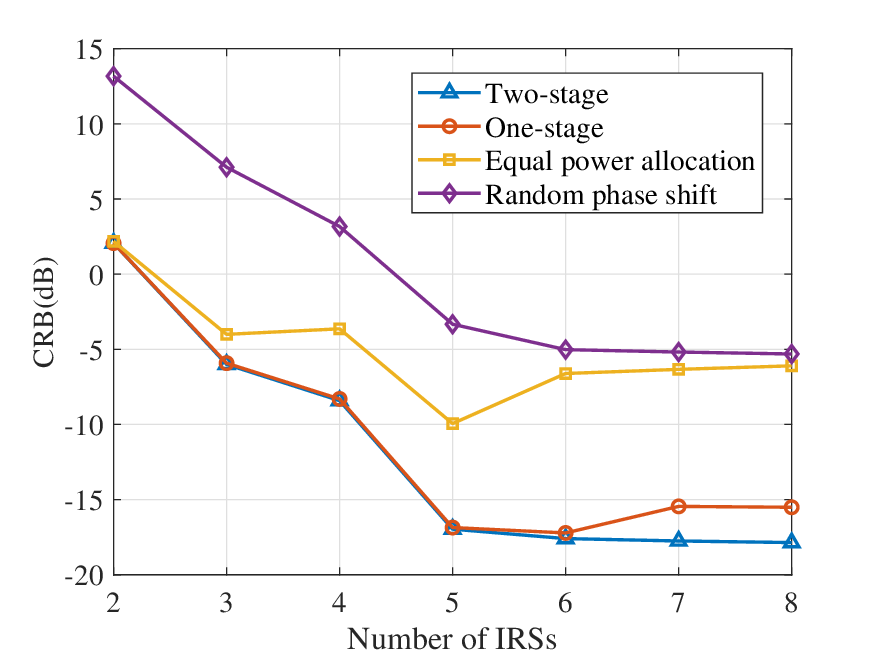}}
	\caption{CRB versus the number of IRSs.}
	\label{s_irs}
\end{figure}

In Fig. \ref{s_irs}, we compare the CRB for location estimation of the single target versus the number of IRSs, $K$. First, it 
is observed that the CRB of both one-stage and equal power allocations schemes fluctuate as the number of IRSs increases.
This is because more spatially distributed IRSs provide more spatial degrees of freedom to localization and signal propagation paths with less path loss. Moreover, there are more sensors on semi-passive IRSs, which provide a larger sample of observations from different angles. The above two reasons explain the positive impact of the increased number of IRSs on localization performance. On the other hand, we did not consider the design of multiple IRS deployments. Therefore, more IRSs will make it difficult to maintain orthogonality among different channels between IRSs and BS for active beamforming at BS. This leads to a reduction in the energy received by the IRSs and thus negatively affects the localization performance. Through the two-stage algorithm, reflective elements on some IRSs will not be activated. This avoids the drawbacks of multiple IRS and causes CRB to monotonically increase as the number of IRS increases.

Second, it is observed from Fig. \ref{s_irs} that when the number of IRSs increases to a certain value, the improvement in localization performance is no longer significant such as when the number of IRSs is raised to 5. This is because the first 5 IRSs already provide a good spatial angle for localizing that target, and only the sensors are functioning in the subsequent added IRSs. Finally, Fig. \ref{s_irs} demonstrates that the one-stage and two-stage algorithms perform comparably when the number of IRSs is small. This is because when the number of antennas is much larger than the number of IRSs, it is simple to achieve channel orthogonality through active beamforming. We will verify it from the case of increasing number of transmitting antennas in the next simulation.

\subsubsection{CRB Versus Number of Transmitting Antennas}
\begin{figure}%[htbp]
\centerline{\includegraphics[width=8cm]{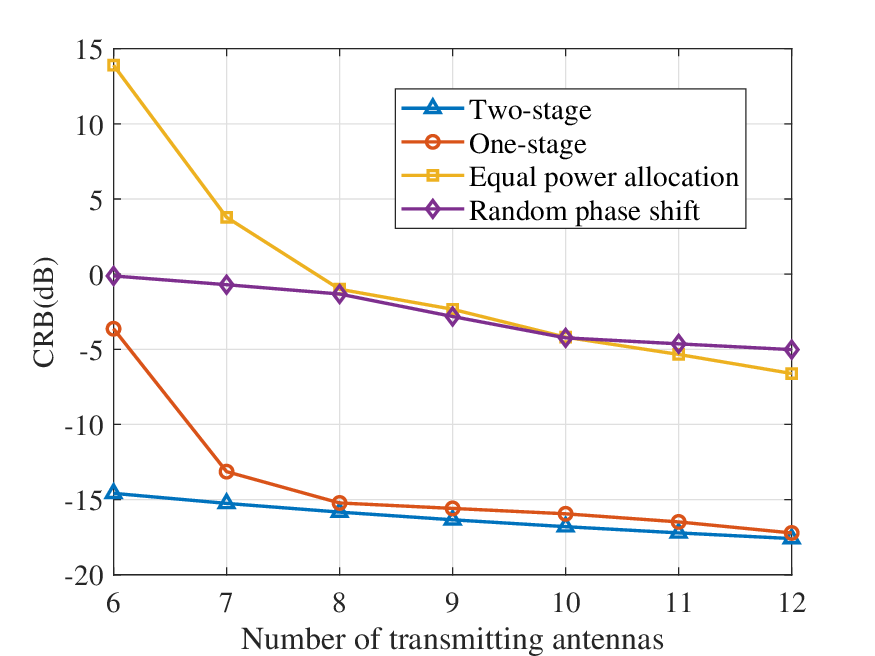}}
	\caption{CRB versus the number of transmitting antennas.}
	\label{single_tr}
\end{figure}

In Fig. \ref{single_tr}, we compare the CRB for location estimation of the single target versus the number of transmitting antennas, $N_t$ when $K=6$. From
Fig. \ref{single_tr}, it is observed that the CRB of all schemes decreases monotonically as $N_t$ increases. When $N_t$ is slightly larger than the number of $K$, increasing $N_t$ improves the performance of the one-stage and equal power allocation schemes significantly while The two-stage scheme achieves the best performance regardless of the number of transmitting antennas. This is because the two-stage scheme can avoid sensing signal interference from redundant IRSs.
The design of the IRS reflection coefficients and the active beamforming  both can bring about a gain of about 15 dB in system performance.
Compared to the one-stage scheme, the advantage of the two-stage scheme is obvious when the values of $N_t$ and $K$ are close.

\subsubsection{CRB Versus Number of Elements or Sensors}
\begin{figure}%[htbp]
\centerline{\includegraphics[width=8cm]{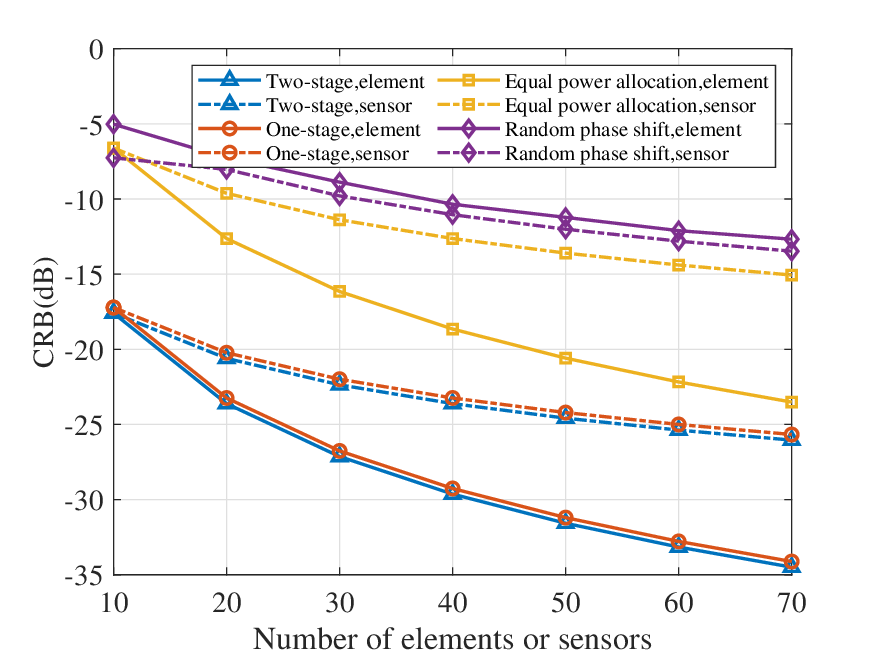}}
	\caption{CRB versus the number of elements or sensors.}
	\label{single_element}
\end{figure}

In Fig. \ref{single_element}, we compare the CRB for location estimation of the single target versus the number of reflective elements, $N$ or sensors, $M$. To make it easier to observe the effect of both $N$ and $M$ on system performance, we set one of them to 10 and change the other from 10 to 70. it is observed that the CRB of all schemes decreases monotonically as $N$ or $M$ increases. However, the order of scaling down of CRB is different with respect to $N$ and $M$. Specifically, CRB scales down with $N$ and $M$ approximately in the order of ${1}/{{{N}^{2}}}$ and ${1}/{M}$ except the random phase shift scheme,  respectively. This is because reflective elements and sensors can achieve beamforming gain of $\mathcal{O}\left( {{N}^{2}} \right)$ and $\mathcal{O}\left( M \right)$,  respectively. However, for random phase shift scheme, reflective elements only achieve beamforming gain of $\mathcal{O}\left( {N} \right)$, which explains that the solid and dashed purple lines almost coincide with each other in Fig. \ref{single_element}.

\subsubsection{CRB Versus Maximum Transmitting Power}
\begin{figure}%[htbp]
\centerline{\includegraphics[width=8cm]{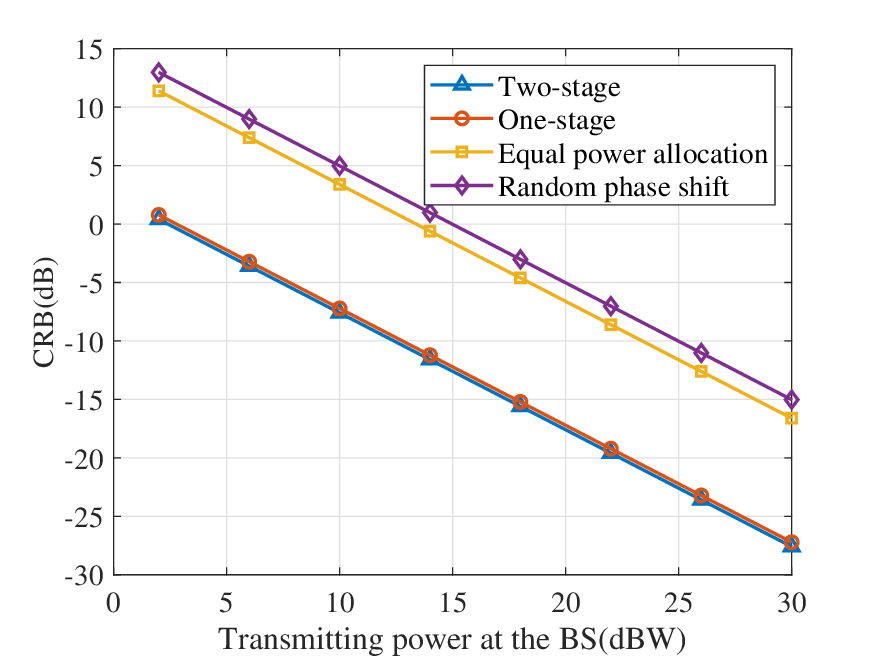}}
	\caption{CRB versus the maximum transmitting power.}
	\label{single_power}
\end{figure}

In Fig. \ref{single_power}, we explored the relationship between CRB and maximum transmitting power, ${{P}_{\max }}$. It is observed that the CRB of all schemes decreases linearly and monotonically with increasing ${{P}_{\max }}$. This is because that Expression for CRB in (\ref{crbq}) is inversely proportional to ${{P}_{\max }}$. All schemes are stable and insensitive to changes in ${{P}_{\max }}$ and the difference between four schemes is constant regardless of ${{P}_{\max }}$.

\subsection{Multi-Target System}
In this subsection, we consider a multi-target system and provide numerical simulation results to verify the effectiveness of the proposed algorithm with the other three schemes.
\subsubsection{CRB Versus Number of IRSs}
\begin{figure}%[htbp]
\centerline{\includegraphics[width=8cm]{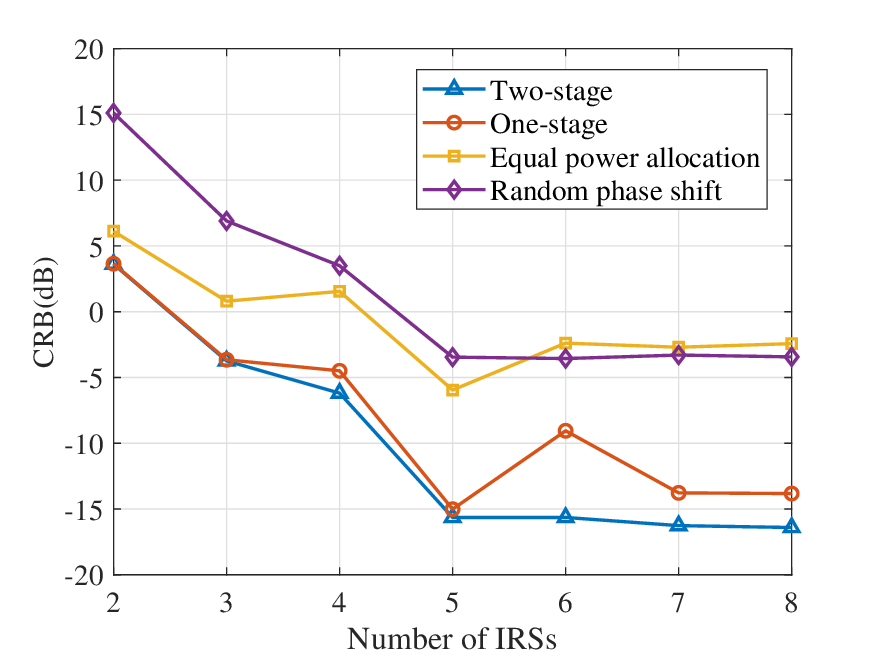}}
	\caption{CRB versus the number of IRSs in multi-target system.}
	\label{multi_irs}
\end{figure}

Fig. \ref{multi_irs} shows the maximum CRB among multi-target versus the number of IRSs with $Q=8$. Like the single-target scenario, the proposed two-stage algorithm alternating-based optimization 
achieves the best performance compared to the other three schemes and can monotonically increase the CRB as the number of IRSs increases.
Compared to the single-target scenario, the performance gap between the proposed algorithm and the random phase design scheme becomes smaller. This is because the random phase design scheme will cause the beam formed by the IRSs to become directionless, and the energy of the sensing signal will be evenly distributed in space. When there are more targets to be located in space, the directionality of the optimal passive beamforming at IRSs will decrease.

\subsubsection{CRB Versus the Number of Targets}
\begin{figure}%[htbp]
\centerline{\includegraphics[width=8cm]{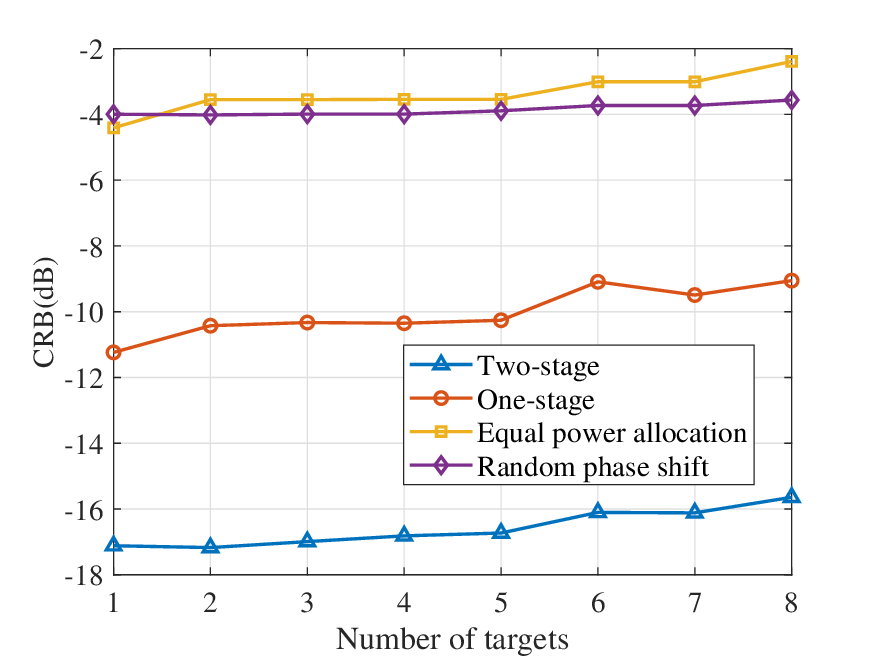}}
	\caption{CRB versus the number of targets in multi-target system.}
	\label{multi_tar}
\end{figure}

Fig. \ref{multi_tar} shows the maximum CRB among multi-target versus the number of targets.
It can be observed that the proposed two-stage algorithm achieves the lowest CRB compared to other benchmark schemes regardless of the number of targets. The CRB of a multi-target system may not increase when some new targets that need to be localized are added, such as the addition of $4$-th and $7$-th targets in Fig. \ref{multi_tar}. This is because we restrict the target to be separable at time and the echo signals from different targets are interference-free with each other. Therefore, when new targets are added to some areas with relatively strong energy coverage, the performance of the whole system will not be degraded, which is similar to the multicast communication system. In addition, we find that the performance of the random phase design scheme remains almost constant as the number of targets increases, which validates the previous conclusions.

\begin{figure*}[!t]
	\normalsize
\begin{equation}\label{der1} 
\begin{aligned}
  & \frac{\partial \log p\left( {{\mathbf{r}}_{\text{IRS}}}|\boldsymbol{\tau}  \right)}{\partial {{\tau }_{{{q}_{1}},{{k}_{1}},{{l}_{1}}}}}= \\ 
 & =\frac{\frac{1}{\sigma _{l}^{2}}\sum\limits_{l=1}^{K}{\int_{T}{{{\left[ {{\mathbf{r}}_{\text{IRS},l}}\left( t \right)-\sum\limits_{q=1}^{Q}{\sum\limits_{k=1}^{K}{{{\mathbf{H}}_{q,k,l}}{{\mathbf{w}}_{k}}{{s}_{k}}\left( t-{{\tau }_{q,k,l}} \right)}} \right]}^{H}}\left[ {{\mathbf{r}}_{\text{IRS},l}}\left( t \right)-\sum\limits_{q=1}^{Q}{\sum\limits_{k=1}^{K}{{{\mathbf{H}}_{m,k,l}}{{\mathbf{w}}_{k}}{{s}_{k}}\left( t-{{\tau }_{q,k,l}} \right)}} \right]dt}}}{\partial {{\tau }_{{{q}_{1}},{{k}_{1}},{{l}_{1}}}}} \\ 
 & =\frac{1}{\sigma _{l}^{2}}\int_{T}{\left\{ {{\left[ {{\mathbf{r}}_{\text{IRS},l}}\left( t \right)-\sum\limits_{q=1}^{Q}{\sum\limits_{k=1}^{K}{{{\mathbf{H}}_{q,k,l}}{{\mathbf{w}}_{k}}{{s}_{k}}\left( t-{{\tau }_{q,k,l}} \right)}} \right]}^{H}}\frac{\partial \left[ {{\mathbf{H}}_{{{q}_{1}},{{k}_{1}},{{l}_{1}}}}{{\mathbf{w}}_{{{k}_{1}}}}{{s}_{{{k}_{1}}}}\left( t-{{\tau }_{{{q}_{1}},{{k}_{1}},{{l}_{1}}}} \right) \right]}{\partial {{\tau }_{{{q}_{1}},{{k}_{1}},{{l}_{1}}}}} \right.} \\ 
 & +\left. \frac{\partial {{\left[ {{\mathbf{H}}_{{{q}_{1}},{{k}_{1}},{{l}_{1}}}}{{\mathbf{w}}_{{{k}_{1}}}}{{s}_{{{k}_{1}}}}\left( t-{{\tau }_{{{q}_{1}},{{k}_{1}},{{l}_{1}}}} \right) \right]}^{H}}}{\partial {{\tau }_{{{q}_{1}},{{k}_{1}},{{l}_{1}}}}}\left[ {{\mathbf{r}}_{\text{IRS},l}}\left( t \right)-\sum\limits_{q=1}^{Q}{\sum\limits_{k=1}^{K}{{{\mathbf{H}}_{q,k,l}}{{\mathbf{w}}_{k}}{{s}_{k}}\left( t-{{\tau }_{q,k,l}} \right)}} \right] \right\}dt, \\ 
\end{aligned}
\end{equation}
	\hrulefill
	\vspace*{1pt}
\end{figure*}

\begin{figure*}[!t]
	\normalsize
\begin{align}\label{der2} 
  & \mathbb{E}\left\{ \frac{{{\partial }^{2}}\log p\left( {{\mathbf{r}}_{\text{IRS}}}|\boldsymbol{\tau} \right)}{\partial {{\tau }_{{{q}_{1}},{{k}_{1}},{{l}_{1}}}}\partial {{\tau }_{{{q}_{2}},{{k}_{2}},{{l}_{2}}}}} \right\}=\frac{1}{\sigma _{n}^{2}}\mathbb{E}\left\{ \int_{T}{\frac{{{\left[ \partial {{\mathbf{H}}_{{{q}_{2}},{{k}_{2}},{{l}_{2}}}}{{\mathbf{w}}_{{{k}_{2}}}}{{s}_{{{k}_{2}}}}\left( t-{{\tau }_{{{q}_{2}},{{k}_{2}},{{l}_{2}}}} \right) \right]}^{H}}}{\partial {{\tau }_{{{q}_{2}},{{k}_{2}},{{l}_{2}}}}}\frac{\partial \left[ {{\mathbf{H}}_{{{q}_{1}},{{k}_{1}},{{l}_{1}}}}{{\mathbf{w}}_{{{k}_{1}}}}{{s}_{{{k}_{1}}}}\left( t-{{\tau }_{{{q}_{1}},{{k}_{1}},{{l}_{1}}}} \right) \right]}{\partial {{\tau }_{{{q}_{1}},{{k}_{1}},{{l}_{1}}}}}} \right. \nonumber\\ 
 & \left. +\frac{{{\left[ \partial {{\mathbf{H}}_{{{q}_{1}},{{k}_{1}},{{l}_{1}}}}{{\mathbf{w}}_{{{k}_{1}}}}{{s}_{{{k}_{1}}}}\left( t-{{\tau }_{{{q}_{1}},{{k}_{1}},{{l}_{1}}}} \right) \right]}^{H}}}{\partial {{\tau }_{{{q}_{1}},{{k}_{1}},{{l}_{1}}}}}\frac{\partial \left[ {{\mathbf{H}}_{{{q}_{2}},{{k}_{2}},{{l}_{2}}}}{{\mathbf{w}}_{k'}}{{s}_{k'}}\left( t-{{\tau }_{{{q}_{2}},{{k}_{2}},{{l}_{2}}}} \right) \right]}{\partial {{\tau }_{{{q}_{2}},{{k}_{2}},{{l}_{2}}}}}dt \right\} \\ 
 & =\left\{ \begin{aligned}
  & 0,{{l}_{1}}\ne {{l}_{2}}\text{ or }{{q}_{1}}\ne {{q}_{2}},\\ 
 & \frac{1}{\sigma _{l}^{2}}\operatorname{Re}\int\limits_{T}{\left[ {{\mathbf{H}}_{{{q}_{1}},{{l}_{1}},k}}\mathbf{H}_{{{q}_{2}},{{l}_{2}},k}^{H}\sum\limits_{k=1}^{K}{\mathbf{w}_{k}^{H}\dot{s}_{k}^{H}\left( t-{{\tau }_{{{q}_{1}},{{l}_{1}},k}} \right)}\sum\limits_{k=1}^{K}{{{\mathbf{w}}_{k}}{{{\dot{s}}}_{k}}\left( t-{{\tau }_{{{q}_{2}},{{l}_{2}},k}} \right)} \right]dt},\text{others}, \nonumber\\ 
\end{aligned} \right.
\end{align}
	\hrulefill
	\vspace*{1pt}
\end{figure*}
\section{Conclusions}
In this paper, we investigate a novel TOA localization-based beamforming optimization problem for multiple IRSs collaborative localization system. Specifically, the active beamforming at the BS, passive beamforming at multiple IRSs, and IRS selection are jointly optimized in the framework of a two-stage algorithm to minimize the estimation CRB of the position of the target in Cartesian coordinates. For single- and multi-target scenarios, we propose ADMM-based and alternating optimization-based algorithms to achieve the best performance compared with other benchmark schemes, respectively. In addition, a number of conclusions that are useful in real system deployments and the gain in system performance with respect to the number of reflective elements and sensors, and the maximum transmitting power are revealed.

\appendices
\section{The Derivation of the FIM in (\ref{fnm1})}
The first derivative of $p\left( {{\mathbf{r}}_{\text{IRS}}}|\boldsymbol{\tau}\right)$ with respect to ${{\tau }_{{{q}_{1}},{{k}_{1}},{{l}_{1}}}}$ is shown in (\ref{der1}), where we have applied the following properties to simplify.
\begin{align}
 \frac{\partial {{\mathbf{r}}_{\text{IRS},l}}\left( t \right)}{\partial {{\tau }_{{{q}_{1}},{{k}_{1}},{{l}_{1}}}}}=0,\frac{\partial s\left( t-{{\tau }_{q,k,l}} \right)}{\partial {{\tau }_{{{q}_{1}},{{k}_{1}},{{l}_{1}}}}}=0,if\text{ }q\ne {{q}_{1}}\text{ or }k\ne {{k}_{1}}. 
\end{align}
Then, we apply the second derivative to (91) with respect to ${{\tau }_{{{q}_{2}},{{k}_{2}},{{l}_{2}}}}$ and result is shown in
(\ref{der2}), where we have applied the following properties to simplify.
\begin{align}
  & \int_{T}{{{{\dot{s}}}_{k}}\left( t-{{\tau }_{{{q}_{1,}},k,l}} \right)\dot{s}_{k}^{H}\left( t-{{\tau }_{{{q}_{2,}},k,l}} \right)}=0,if\text{ }{{q}_{1,}}\ne {{q}_{2}}, \\ 
 & \mathbb{E}\left\{ {{\mathbf{r}}_{\text{IRS},k}}\left( t \right)-\sum\limits_{q=1}^{Q}{\sum\limits_{k=1}^{K}{{{\mathbf{H}}_{q,k,l}}{{\mathbf{w}}_{k}}{{s}_{k}}\left( t-{{\tau }_{q,k,l}} \right)}} \right\}=0.  
\end{align}

\section{Details of Algorithm}
The Lagrangian function of (P2.5) can be formulated as
\begin{align}
& L\left( \mathbf{z},{{\mu }_{0}},\boldsymbol{\mu } \right)=\frac{\rho }{2}\left\| \mathbf{p}-\mathbf{z} \right\|_{2}^{2}-\beta {{\mathbf{z}}^{T}}\mathbf{z}+{{\boldsymbol{\lambda }}^{T}}\left( \mathbf{p}-\mathbf{z} \right) \nonumber\\ 
& +{{\mu }_{0}}\left( {{\mathbf{h}}^{T}}\mathbf{z}-{{P}_{\max }} \right)+{{\boldsymbol{\mu }}^{T}}\mathbf{z}, 
\end{align}
where ${\mu }_{0}$ and ${{\boldsymbol{\mu }}}$ are Lagrange multipliers associated with constraint \text{(\ref{zcon1})} and \text{(\ref{zcon2})},  respectively. According to KKT conditions, we have
\begin{gather}
\frac{\partial L}{\partial {{z}_{i}}}=\left( \rho -2\beta  \right){{z}_{i}}-\left( \rho {{p}_{i}}+{{\lambda }_{i}} \right)+{{\mu }_{0}}{{h}_{i}}+{{\mu }_{i}}=0,\forall i, \label{kkt1}\\ 
{{\mathbf{h}}^{T}}\mathbf{z}-{{P}_{\max }}=0, \label{kkt2}\\ 
{{z}_{i}}\ge 0,\forall i, \label{kkt3}\\ 
{{\mu }_{i}}\ge 0,\forall i, \label{kkt4}\\ 
{{\mu }_{i}}{{z}_{i}}=0,\forall i.\label{kkt5} 
\end{gather}
Based on (\ref{kkt1}) and (\ref{kkt4}), we have
\begin{align}
{{\mu }_{i}}=\left( \rho {{p}_{i}}+{{\lambda }_{i}} \right)-\left( \rho -2\beta  \right){{z}_{i}}-{{\mu }_{0}}{{h}_{i}}\ge 0,\forall i.
\end{align}
According to complementary relaxation condition (\ref{kkt5}), 
\begin{align}
\left[ \left( \rho {{p}_{i}}+{{\lambda }_{i}} \right)-\left( \rho -2\beta  \right){{z}_{i}}-{{\mu }_{0}}{{h}_{i}} \right]{{z}_{i}}=0,\forall i.
\end{align}
Then, we can discuss the solution based on the value of ${\mu }_{0}$

\textbf{Case 1}: ${{\mu }_{0}}\ge 
\frac{\rho {{p}_{i}}+{{\lambda }_{i}}}{{{h}_{i}}}$

It is obvious that $\left( \rho {{p}_{i}}+{{\lambda }_{i}} \right)-\left( \rho -2\beta  \right){{z}_{i}}-{{\mu }_{0}}{{h}_{i}}\ge 0$ because ${{z}_{i}}\ge 0$ and $\rho >2\beta $. Therefore, the 
sufficient and necessary conditions
of (\ref{kkt5}) is ${{z}_{i}}=0$.

\textbf{Case 2}: ${{\mu }_{0}}<\frac{\rho {{p}_{i}}+{{\lambda }_{i}}}{{{h}_{i}}}$

If ${{z}_{i}}=0$, (\ref{kkt5}) must not hold. Therefore, 
\begin{align}\label{zi}
{{z}_{i}}=\frac{\rho {{p}_{i}}+{{\lambda }_{i}}-{{\mu }_{0}}{{h}_{i}}}{\rho -2\beta }, 
\end{align}
which ensures 
(\ref{kkt5}) holds.

The final problem is to determine the value of ${{\mu }_{0}}$. Firstly, we are unable to determine the numerical relationship between ${{\mu }_{0}}$ and $\frac{\rho {{p}_{i}}+{{\lambda }_{i}}}{{{h}_{i}}}$,$\forall i$. Therefore we assume that ${{\mu }_{0}}<\frac{\rho {{p}_{i}}+{{\lambda }_{i}}}{{{h}_{i}}},\forall i$.
Recalling (\ref{kkt2}) and (\ref{zi}), we
have
\begin{align}\label{miu0}
{{\mu }_{0}}=\frac{-\left( \rho -2\beta  \right){{P}_{\max }}+\sum\limits_{i=1}^{K}{\left( \rho {{p}_{i}}{{h}_{i}}+{{\lambda }_{i}}{{h}_{i}} \right)}}{\sum\limits_{i=1}^{K}{h_{i}^{2}}}.
\end{align}
Substituting (\ref{miu0}) into (\ref{zi}), we can obtain ${{z}_{i}}$. If, according to the calculations, $\exists i',{{z}_{i'}}\le 0$, we set ${{z}_{i'}}=0,\forall i'$ and update ${{\mu }_{0}}$ and recalculate ${{z}_{i}}$ until ${{z}_{i}}\ge 0,\forall i$.

\bibliographystyle{IEEEtran}
\bibliography{reference}
\end{document}